\documentclass[iop]{emulateapj}
\usepackage{natbib}
\usepackage{amsmath}
\usepackage{graphicx}
\usepackage{subfigure}
\usepackage{float}
\usepackage{epsfig}
\usepackage{booktabs}
\usepackage{CJK}

\shorttitle{Pebble Evolution and Formation Timescales}
\begin{document}
\begin{CJK*}{UTF8}{gkai}

\title{Inside-Out Planet Formation. IV. Pebble Evolution and Planet Formation Timescales}

\author{Xiao Hu (胡晓)\altaffilmark{1,2}, Jonathan C. Tan\altaffilmark{1,3}, Zhaohuan Zhu (朱照寰)\altaffilmark{2}, Sourav Chatterjee\altaffilmark{4}, Tilman Birnstiel\altaffilmark{5}, 
\\
Andrew N. Youdin\altaffilmark{6}, Subhanjoy Mohanty\altaffilmark{7}}
\altaffiltext{1}{Department of Astronomy, University of Florida, Gainesville, FL 32611}
\altaffiltext{2}{Department of Physics and Astronomy, University of Nevada, Las Vegas, 4505 South Maryland Parkway, Las Vegas, NV 89154}
\altaffiltext{3}{Department of Physics, University of Florida, Gainesville, FL 32611}
\altaffiltext{4}{Center for Interdisciplinary Exploration and Research in Astrophysics (CIERA), Physics and Astronomy, Northwestern University, Evanston, IL 60208}
\altaffiltext{5}{University Observatory, Faculty of Physics, Ludwig-Maximilians-Universität München, Scheinerstr. 1, 81679 Munich, Germany}
\altaffiltext{6}{Department of Astronomy and Steward Observatory, University of Arizona, 933 North Cherry Avenue, Tucson, AZ 85721}
\altaffiltext{7}{Imperial College London, 1010 Blackett Lab, Prince Consort Rd., London SW7 2AZ, UK}
\begin{abstract}
Systems with tightly-packed inner planets (STIPs) are very common.
\citet{CT14} proposed Inside-Out Planet Formation (IOPF), an {\it in
  situ} formation theory, to explain these planets. IOPF involves
sequential planet formation from pebble-rich rings that are fed from
the outer disk and trapped at the pressure maximum associated with the
dead zone inner boundary (DZIB). Planet masses are set by their
ability to open a gap and cause the DZIB to retreat outwards.
We present models for the disk density and temperature structures that
are relevant to the conditions of IOPF. For a wide range of DZIB
conditions, we evaluate the gap opening masses of planets in these
disks that are expected to lead to truncation of pebble accretion onto
the forming planet. We then consider the evolution of dust and pebbles
in the disk, estimating that pebbles typically grow to sizes of a few
cm during their radial drift from several tens of AU to the inner,
$\lesssim1\:$AU-scale disk. A large fraction of the accretion flux of
solids is expected to be in such pebbles. This allows us to estimate
the timescales for individual planet formation and entire planetary
system formation in the IOPF scenario. We find that to produce
realistic STIPs within reasonable timescales similar to disk lifetimes
requires disk accretion rates of $\sim10^{-9}\:M_\odot\:{\rm yr}^{-1}$
and relatively low viscosity conditions in the DZIB region, i.e.,
Shakura-Sunyaev parameter of $\alpha\sim10^{-4}$.
\end{abstract}

\section{Introduction}

More than 4000 exoplanet candidates
\citep[e.g.,][]{2015ApJS..217...31M} have been discovered by {\it
  Kepler} since 2009, a significant percentage ($\gtrsim 30\%$) of
which are in systems with tightly-packed inner planets (hereafter
STIPs). STIPs are very different from our Solar System: they typically
contain three or more detected planets of radii $\sim1-10\:R_\oplus$
and with periods less than 100 days \citep{2012ApJ...761...92F}, i.e.,
with orbital radii of $\lesssim 0.3$~AU.

One way to form STIPs may be via inward migration of planets formed in
the outer disk \citep[e.g.,][]{2012ARA&A..50..211K,
  2013A&A...553L...2C,2014IAUS..299..360C}. However, it can be quite
difficult to concentrate planets close to the host star to the degree
of observed STIPs \citep{2010MNRAS.401.1691M}. Also migrating planets
tend to become trapped in orbits of low order mean motion resonances,
which is not a particular feature of STIPs
\citep{2014prpl.conf..667B,2014ApJ.790.146}. This lack of resonant
pile-up has motivated studies of mechanisms that cause a lower
efficiency of resonance trapping \citep{2014AJ....147...32G} or
breaking of resonance by later dynamical interaction with, e.g., a
planetesimal disk \citep[][]{2015ApJ...803...33C}, by the
  unstable nature of compact resonant chains
  \citep{2017MNRAS.470.1750I}, or by magnetospheric rebound from the
host star during gas disk dispersal \citep{2017arXiv170202059L}.

As an alternative to inward migration, {\it in situ} formation
scenarios \citep{2012ApJ.751.158, 2013ApJ.775.53}
have considered a group of protoplanets that are initially distributed
inside 1 AU with high concentration. After following dynamical
evolution for 10 Myr, collisional oligarchic growth can lead to
systems quite similar to STIPs. However, how this initial condition
was established was not well studied in these works and may be subject
to some difficulties, such as early triggering of gravitational
instability \citep{2014MNRAS.440L..11R,2014ApJ...795L..15S} 
\citep[see also,][]{2013MNRAS.431.3444C}.  In addition, simulations that account for
the effects of gas on protoplanet migration during the oligarchic
growth phase find that systems are produced with planet masses
decreasing steeply with orbital radius, which is different from the
relatively flat scaling seen in STIPs \citep{2015A&A...578A..36O}.

The Inside-Out Planet Formation (IOPF) scenario proposed by
\citep[][hereafter  Paper I]{CT14} is a new type of {\it in situ}
formation model. It starts with pebble delivery to the disk midplane
transition region between an innermost magnetorotational instability
(MRI) active zone and a nonactive ``dead zone,'' where there is a
local pressure maximum. The pebbles are trapped here and build up in a
ring, which then forms a first protoplanet, perhaps involving a
variety of processes including streaming \citep{2005ApJ...620..459Y},
gravitational \citep{1964ApJ...139.1217T} and/or Rossby wave
\citep{2006A&A...446L..13V} instabilities.  The protoplanet is
expected to continue its growth, especially by pebble accretion, until
it becomes massive enough to open a gap in the disk. This gap, which
may only be a partial gap with a modest reduction in gas mass surface
density, pushes the pressure maximum outwards by at least a few Hill
radii from the first planet, thus creating a new pebble trap, and thus
location of second planet formation \citep[][hereafter Paper
  III]{2016ApJ...816...19H} \citep[see also][]{2014A&A...572A..35L}.
Furthermore, MRI-activation in this gap region beyond the first planet
may induce enhanced viscosity and thus further outward retreat of the
dead zone inner boundary (DZIB) pressure trap, leading to larger
separations to the orbit of the second planet.

The pebble delivery rate is crucial for setting the timescale of
pebble ring and planet formation in the IOPF scenario. To calculate
this, we need to model the evolution of dust and pebbles in the
protoplanetary disk and determine if the majority of solid material,
which in the outer disk is initially in small dust grains, become
incorporated into pebbles by the time it reaches the inner
region. The IOPF model requires dust grains to grow into pebbles as
smaller grains are well coupled to the gas and will not be trapped at
a local pressure maximum \citep[e.g.,][]{2012ApJ...755....6Z}.  The size
distribution of pebbles is also important for models of planetesimal
formation, e.g., via the streaming instability. This may be triggered
in both the pebble ring as a first step for {\it in situ} planet
formation, or in the outer disk as a potential first step for outer
planet formation that may eventually truncate pebble delivery to the
inner disk. Finally, modeling of dust and pebble evolution is a
necessary step for then providing observational predictions of the
emission from the disk during various phases of IOPF.

There have been a number of previous studies of dust and pebble
evolution in global disk models, including radial drift and particle
interaction \citep[e.g.,][]{Brauer2008a,Birnstiel2010}. 
\citet{Birnstiel2012} proposed a two population model that divides
dust into one smaller fixed size group and one larger variable size
group that represents grain growth. \citet{2015arXiv151202414S}
proposed a simplified pebble-pebble interaction scenario by reducing
the full Smoluchowski equation to a single sized evolution equation,
yielding pebble fluxes and typical sizes in disks on scales down to
$\sim$1 AU. A Lagrangian dust evolution model has been investigated by
\citet{Krijt2016}, which tracks batches of particles in the disk as
they drift radially and grow through collisions.

In this paper, we present a new pebble-dust evolution model and apply
it specifically to calculate the timescales needed to build up a STIP
forming via IOPF, i.e., limited by the rate of pebble delivery. 
In \S\ref{S:disk}, we review the equations that govern the disk models
that are relevant to IOPF, including presenting new disk structure
models that include realistic opacities and the transition from an
``active,'' accretion-powered inner disk to a ``passive,''
stellar-heated outer disk. In \S\ref{S:solid flux}, we discuss a
simple estimate of the timescale of first, so-called ``Vulcan'' planet
formation in steady accretion rate disks, including a reanalysis of
the gap-opening mass criterion that was studied in Paper III. We also
estimate the size of the region of the disk that is needed to supply
the material for this planet. In \S\ref{S:single}, we model the radial
drift and sweep up growth of a single pebble in our model disks. In
\S\ref{S:model}, we introduce our full pebble evolution model and
numerical results for different disk parameters.  In \S\ref{S:STIP} we
outline the assembly of a STIPs-like system with pebble flux and
planet mass criterion derived from earlier sections.  We discuss the
implications of our results and summarize our conclusions in
\S\ref{S:conclusions}.

\section{Disk Structure}\label{S:disk}

The structure of a steady accretion disk can be specified with a given
accretion rate $\dot{m}$, \citet{1973A&A....24..337S} viscosity
parameter $\alpha$, stellar mass $m_*$, and midplane temperature
versus radius relation
$T(r)$ (in particular specifying whether the energy source of the disk
is heating mostly by its own accretion, i.e., an ``active'' disk, or
whether it is heating mostly by stellar irradiation, i.e., a
``passive'' disk):
\begin{eqnarray}
T & = & T(r)\\
c_s & = & (\gamma k_B T/\mu)^{1/2}\\
h/r & = & c_s/v_K\\
\nu & = &\alpha c_s h\\
\Sigma_g & = &\dot{m}/(3\pi\nu)\\
\rho & = &\Sigma_g/(h\sqrt{2\pi}),
\end{eqnarray} 
where $c_s$ is the midplane sound speed, $\gamma$ is the power law
exponent of the barotropic equation of state $P=K\rho^\gamma$, $k_B$
is Boltzmann's constant,
$\mu=2.33m_{\rm{H}}=3.90\times10^{-24}\:\rm{g}$ is the mean particle
mass (with fiducial value set by assuming $n_{\rm He}=0.2n_{\rm H2}$),
$h$ is the disk vertical scaleheight, $v_K$ is the local Keplerian
speed, $\nu$ is the viscosity, $\alpha \equiv 10^{-4}\alpha_{-4}$ is
the Shakura-Sunyaev dimensionless viscosity parameter (with fiducial
value set for dead zone conditions, but we will consider other values
also), $\Sigma_g$ is the gas mass surface density, $\dot{m}\equiv
\dot{m}_{-9} 10^{-9}\:M_\odot\:{\rm yr}^{-1}$ is the accretion rate
(with fiducial normalization set by observed accretion rates of
transition disks, but again we will consider a range of values; see
 Paper I) and $\rho$ is midplane gas density. Only equations (1) and (2)
depend on the energy source and equation of state, while the rest are
general equations followed by all viscous disks.

The solution for the structure of an active accretion disk is achieved
by balancing viscous thermal dissipation with vertical radiative
cooling (per unit area from one face of the disk):
\begin{equation}
\frac{\nu\Sigma_g r^2}{2}\left(\frac{d\Omega}{dr}\right)^2=\frac{16\sigma_{\rm SB}}{3\Sigma_g\kappa}T^4
\end{equation}
This generally applies in the inner disk. In this case the temperature equation is
(Paper I):
\begin{eqnarray}
T&=&\frac{3^{1/5}}{2^{7/5}\pi^{2/5}}\left(\frac{\mu}{\gamma k_B}\right)^{1/5}\left(\frac{\kappa}{\sigma_{\rm SB}}\right)^{{1}/{5}} \nonumber \\
&\times& \alpha^{-1/5}\left(Gm_*\right)^{3/10}\left(f_r\dot{m}\right)^{2/5}r^{-{9}/{10}},\\
&\rightarrow&290 \gamma_{1.4}^{-1/5}\kappa_{10}^{1/5}\alpha_{-4}^{-1/5}m_{*,1}^{3/10}(f_r\dot{m}_{-9})^{2/5}r_{\rm AU}^{-9/10}\: \rm K, \nonumber
\end{eqnarray}
and gas mass surface density is given by:
\begin{eqnarray}
\label{eq:sigma_g}
\Sigma_g&=&\frac{2^{7/5}}{3^{6/5}\pi^{3/5}}\left(\frac{\mu}{\gamma k_B}\right)^{4/5}\left(\frac{\kappa}{\sigma_{\rm SB}}\right)^{-{1}/{5}} \nonumber \\
&\times& \alpha^{-4/5}\left(Gm_*\right)^{1/5}\left(f_r\dot{m}\right)^{3/5}r^{-{3}/{5}},\\
&\rightarrow& 812\gamma_{1.4}^{-4/5}\kappa_{10}^{-1/5}\alpha_{-4}^{-4/5}m_{*,1}^{1/5}(f_r\dot{m}_{-9})^{3/5}{r_{\rm AU}}^{-3/5}\: \rm g\:cm^{-2},\nonumber
\end{eqnarray}
where $\gamma\equiv1.4\gamma_{1.4}$ with fiducial normalization to a
value of 1.4 for $\rm H_2$ with rotational modes excited, $\kappa
\equiv \kappa_{10}10\ {\rm cm^2\:g^{-1}}$ is the disk mean opacity
\citep[with fiducial normalization here appropriate for inner disk
conditions; but, note we will more generally use tabulated opacities
from][]{zhu2009opac},
$\sigma_{\rm SB}$ is the Stefan-Boltzmann constant, $m_*\equiv
m_{*,1}M_\odot$ is the stellar mass with fiducial normalization of
$1\:M_\odot$, $f_r \equiv1-\sqrt{r_*/r}$, $r_*$ is the stellar radius,
and $r_{\rm AU}\equiv r/(1\:{\rm AU})$. The two parameters with some
of the largest uncertainties are $\alpha$ and $\dot{m}$, so we shall
consider the effects of varying their fiducial values.

In the outer disk, stellar irradiation is expected to become the
dominant energy source compared to local viscous accretion
heating. Following \citet{CG97}, the temperature structure of a
stellar irradiated, passive disk follows $T\propto r^{-3/7}$, which is
much shallower than the $T\propto r^{-9/10}$ relation of the active
disk.

To derive $T(r)$ in the passive disk regime, the following equations
are solved iteratively:
\begin{eqnarray}
T&=&(\beta/4)^{1/4}(r_*/r)^{1/2}T_{\rm *,eff};\\
\beta&=&0.4r_*/r+\eta H/r.
\end{eqnarray}
Here $\beta$ is the ``grazing angle,'' i.e., the angle of incidence of
stellar radiation at the disk surface, $T_{\rm *,eff}$ is the surface,
photospheric temperature of the star, $\eta=2/7$ is the flaring index,
i.e., $\eta=d(h/r)/dr$, and $H$ is the height of the disk photosphere
above the midplane. Usually, $H=4h$ is a good approximation in most
parts of the disk \citep{CG97}. If we choose a solar mass T-Tauri star
with temperature $T_{\rm *,eff}=4500\:K$ and radius $r_*=3.0\:
R_\odot=0.014 \rm AU$, we have the disk temperature versus radius
equation
\begin{eqnarray}
T & = & 172 r_{\rm AU}^{-3/7}\: \rm K,
\end{eqnarray}
which is independent of $\alpha$ and $\dot{m}$.

For our full numerical solutions, we set up a hybrid disk that is
solved first in its inner region as an active disk, and then calculate
a transition radius where the contribution from stellar irradiation
becomes equal to that due to accretion heating. Beyond this transition
radius, we scale the temperature in the manner appropriate for the
passive disk regime, i.e., in particular, we compare the midplane
temperatures of the active and passive disk models and choose the
hotter one as the actual value. 

\begin{figure*}
\centering
\includegraphics[width=\textwidth]{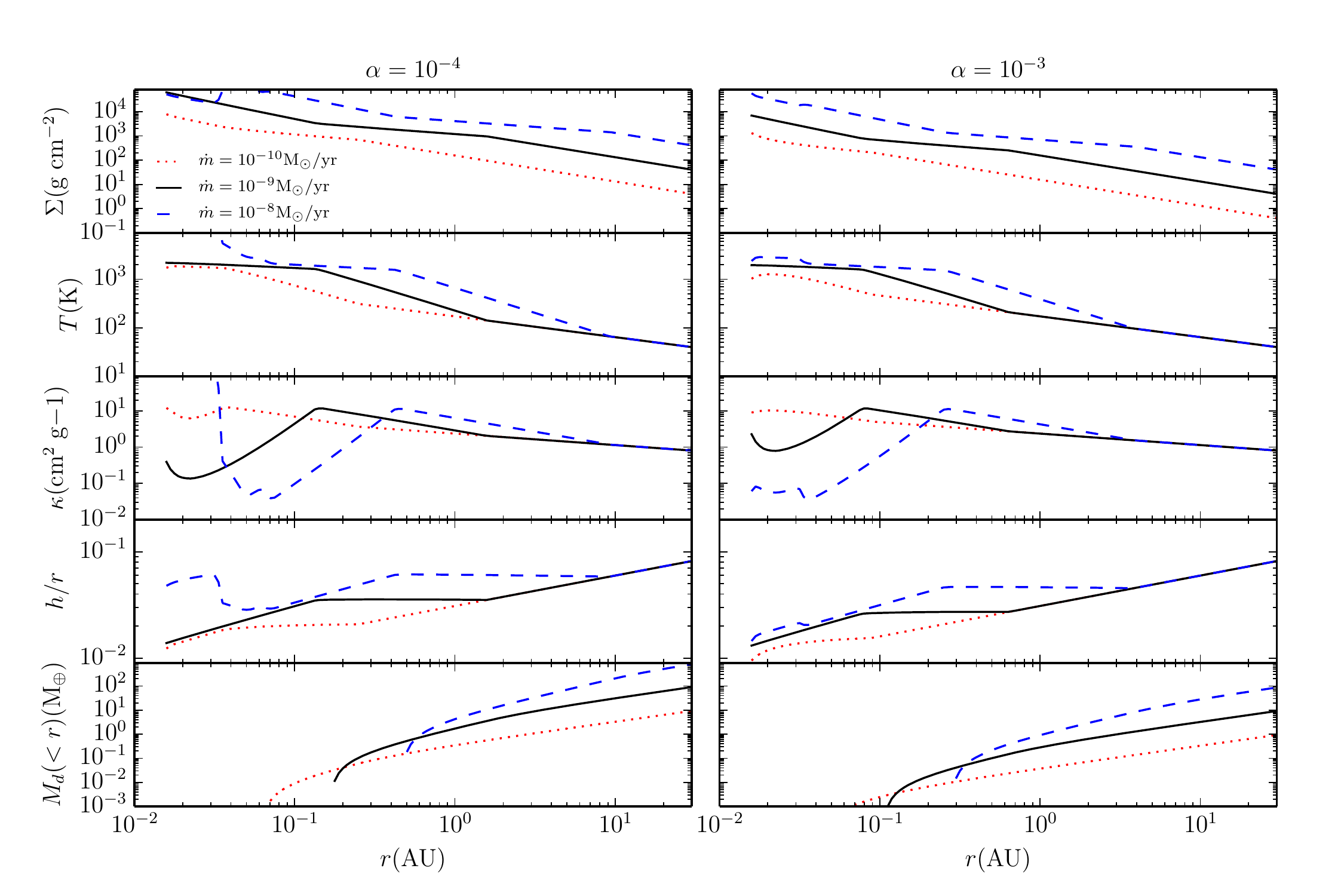}
\caption{
Structural profiles of IOPF disk models used in this paper, assuming
$\alpha=10^{-4}$ (left column) and $\alpha=10^{-3}$ (right column),
for accretion rates $10^{-10}$ (red dotted), $10^{-9}$ (black solid)
and $10^{-8}\:M_\odot\:{\rm yr}^{-1}$ (blue dashed). All models are
for a one solar mass central star. From top to bottom, the rows show:
gas mass surface density ($\Sigma_g$); midplane temperature ($T$);
midplane opacity ($\kappa$) (assumed to be vertically constant); disk
aspect ratio ($h/r$); enclosed mass in solids ($M_d(<r)$), i.e.,
initially assumed to be dust, summed in the disk from $r_{\rm 1200K}$
out to radius $r$ for a solid to gas mass ratio of $f_s=0.01$. Note
that the model assumption of disk vertical optical depth being
$\gtrsim1$ breaks down in the cases with $\alpha=10^{-3}$ and $\dot{m}
\leq 10^{-9}\:M_\odot\:{\rm yr}^{-1}$ in the outer disk, e.g., for
$\dot{m} = 10^{-9}\:M_\odot\:{\rm yr}^{-1}$ this occurs at $r>8.5$~AU.
}
\label{fig:disk}
\end{figure*}

The results of these constant $\alpha=10^{-4}$ (the fiducial value)
and $10^{-3}$ and constant $\dot{m}=10^{-10}$, $10^{-9}$ and
$10^{-8}\:M_\odot\:{\rm yr}^{-1}$ disk models are shown in
Figure~\ref{fig:disk}. Such values of $\alpha$ have been inferred in
the inner regions of simulated dead zones \citep{2010A&A...515A..70D}. Whether such dead zone conditions and values of $\alpha$
continue to apply in larger scale regions of the disk remains an open
question \citep[e.g.,][]{2013ApJ.764.65}, however, for simplicity we will
assume that they do and calculate the structure of the disk out to
30~AU. The range of accretion rates that we consider is guided by
observations of transition disks \citep[e.g.,][]{manara}. Another
risk of applying constant $\alpha$ and stable accreting disk to larger
radius is overestimating the disk mass. In our fiducial case, as seen
in Figure~\ref{fig:disk}, the dust mass within 30 AU is about 100 $M_\oplus$,
while recent observations on dust continuum \citep{2016A&A...592A.126V} give
an average around 40 $M_\oplus$ for most transitional disks. This factor
of 2 should be considered when analyzing the simulation result. 
The active to passive transition radius for the fiducial disk
($\dot{m} = 10^{-9}M_\odot\:{\rm yr}^{-1}, \alpha=10^{-4}$) is
1.6~AU. We note that the model assumption of disk vertical optical
depth being $\gtrsim1$ breaks down in the cases with $\alpha=10^{-3}$
and $\dot{m} \leq 10^{-9}\:M_\odot\:{\rm yr}^{-1}$ in the outer disk,
e.g., for $\dot{m} = 10^{-9}\:M_\odot\:{\rm yr}^{-1}$ this occurs for
$r>8.5$~AU. Note also that an ``inner transition'' in disk structure
at 0.14~AU results from the drop in opacity due to dust
sublimation. However, in the context of IOPF, we expect $\alpha$ to
also rise dramatically when $T\sim1200\:$K (this location is referred
as $r_{\rm 1200K}$ hereafter) as the MRI is activated by thermal
ionization of alkali metals. This change in $\alpha$ has not been
included in the models shown in Fig.~\ref{fig:disk} since our main
focus here is modeling the region that is the supply reservoir for
solid material delivered to $\sim$0.1~AU locations.  Still, we have
extended the disk models within 0.1~AU, since for the low accretion
rate case ($\dot{m} = 10^{-10}M_\odot\:{\rm yr}^{-1}$), $r_{\rm 1200K}
< 0.1\: \rm AU$.

In fact, there are some close-in planets with orbits within 0.1~AU,
e.g., {\it Kepler}-10b \citep{Kepler10} is a super-Earth planet
orbiting a solar type star at 0.016~AU. For such planets to form {\it
  in situ} places constraints on IOPF models. For example, such
systems would need to form in a disk with relatively low accretion
rate that has a small value of $r_{\rm 1200K}$ (see below).

\section{``Vulcan'' planet formation}
\label{S:solid flux}

Our goal in this section is to obtain a simple estimate for the
formation time and the size of the supply reservoir of the first,
i.e., innermost planet to form in the IOPF scenario 
\citep[][hereafter Paper II]{CT15}.
To do this requires knowing the mass of such planets and the mass
flux of pebbles. Here we will assume a steady accretion rate with most
mass flux of solids being in the form of pebbles. The validity of this
assumption will be explored in the following sections of the paper.

\subsection{Location}

The radial location of Vulcan planet formation is assumed to be set by
where the temperature reaches $\sim$1,200~K that leads to thermal
ionization of alkali metals, especially Na and K. This location varies
depending on the accretion rate and other disk properties.
In an active accretion disk, which is the relevant regime for the
optically thick inner disk, this location is (Paper I):
\begin{equation}
\label{eq:rinner}
r_{1200{\rm K}}=0.218\phi_{\rm DZIB}\gamma_{1.4}^{-2/9}\kappa_{10}^{2/9}\alpha_{-4}^{-2/9}m_{*,1}^{1/3}(f_r\dot{m}_{-9})^{4/9}\:{\rm AU},
\end{equation}
where $\phi_{\rm DZIB}$ is a dimensionless parameter of order unity to
account for potential differences from the pure viscous disk model,
e.g., due to the effects of extra energy extraction on midplane
temperature by a disk wind. Paper I and II adopted a fiducial value of
$\phi_{\rm DZIB}$ of 0.5. In Paper III, we used 0.1 AU as the typical
location of Vulcan planets for hydrodynamic simulations of gap
opening, where $\alpha=10^{-3}$, $\dot{m}=10^{-9}\:M_\odot\:{\rm
  yr}^{-1}$, which corresponds to $\phi_{\rm DZIB} = 0.76$.

\subsection{Mass}

The mass of the Vulcan planet (and all planets forming via IOPF) is
assumed to be set by gap opening, which then leads to displacement of
the local pressure maximum away from the planet, to a larger radius in
the disk, and thus the truncation of pebble accretion. Paper III
investigated this process with 2D hydrodynamic simulations for the
case of the fiducial accretion rate ($10^{-9}\:M_\odot\:{\rm
  yr}^{-1}$), a value of $\alpha=10^{-3}$ for the inner dead zone that
then rises to $\alpha=10^{-2}$ in the MRI-active region, and for the
planet set at a fixed location of 0.1~AU from a $1\:M_\odot$ star. The
mass of the planet leading to the first significant displacement of
the pressure maximum was assessed relative to the viscous thermal
criterion gap opening mass of \citet{Lin1993gap}:
\begin{eqnarray}
M_{G}&=&\phi_G\frac{40\nu m_*}{r^2 \Omega_K}\label{eq:MG}\\
&=&40\phi_G \left(\frac{3}{128}\right)^{1/5}\pi^{-{2}/{5}}\left(\frac{\mu}{k_B}\right)^{-{4}/{5}}\gamma^{{4}/{5}}\sigma_{\rm SB}^{-{1}/{5}}
\nonumber\\
&\times&\alpha^{{4/5}}G^{-{7/10}}m_*^{{3}/{10}}\kappa^{1/5}\left(f_r \dot{m}\right)^{2/5}r^{{1}/{10}}\nonumber\\
&\rightarrow&11.2\phi_{G}\gamma_{1.4}^{4/5}\kappa_{10}^{1/5}\alpha_{-3}^{4/5}m_{*,1}^{3/10}(f_r\dot{m}_{-9})^{2/5} r_{\rm 0.1AU}^{1/10}\:M_\oplus\nonumber\\
&\rightarrow&1.77\phi_{G}\gamma_{1.4}^{4/5}\kappa_{10}^{1/5}\alpha_{-4}^{4/5}m_{*,1}^{3/10}(f_r\dot{m}_{-9})^{2/5} r_{\rm 0.1AU}^{1/10}\:M_\oplus.\nonumber
\end{eqnarray}
Here $r_{\rm 0.1AU}\equiv r/(0.1\:{\rm AU})$. 
The last two numerical evaluations of this mass illustrate its
relatively strong dependence on $\alpha$. Paper III found that, for
the $\alpha=10^{-3}$ case, the critical pressure-displacement gap
opening mass was about 50\% of the viscous-thermal gap opening mass
(i.e., this fraction was expressed as $\phi_G\simeq0.5$). Note that
the mass evaluated in equation~(\ref{eq:MG}) is assessed only from the
viscous criterion, but, for the conditions with $\alpha=10^{-3}$, is
consistent with the thermal criterion also, i.e., the Hill radius of
the planet is similar in size to the disk scale height. However, as we
see now, this condition breaks down when $\alpha=10^{-4}$, as the
planet mass is much lower.

To better capture the behavior of gap opening for more general
conditions in which the thermal criterion also begins to limit gap
opening, we now improve our treatment by adopting the
 expression of \citet{Duffell2015} for the
gap opening mass (same scalings of $h/r$ and $\alpha$ is
also presented in \citet{2013ApJ...768..143Z} and \citet{2014ApJ...782...88F}):
\begin{eqnarray}
\label{eq:mg1}
M_{\rm G,D}&=&\phi_{\rm G,D}\sqrt{3\pi}m_*\alpha^{1/2}\left(h/r\right)^{5/2}\\
&=&\phi_{\rm G,D}\frac{3^{3/4}}{2^{7/4}}\left(\frac{\mu}{\gamma k_B}\right)^{-1}\left(\frac{\kappa}{\sigma_{\rm SB}}\right)^{{1}/{4}} \alpha^{1/4}\nonumber\\
&\times&G^{-7/8}m_*^{1/8}\left(f_r\dot{m}\right)^{1/2}r^{{1}/{8}},\nonumber\\
&\rightarrow& 2.68\phi_{\rm G,D} \gamma_{1.4}\kappa_{10}^{1/4}\alpha_{-4}^{1/4}m_{*,1}^{1/8}(f_r\dot{m}_{-9})^{1/2}r_{\rm 0.1 AU}^{1/8}\: \rm M_\oplus \nonumber
\end{eqnarray}
where $\phi_{\rm G,D}$ is a dimensionless parameter of order unity
(note this absorbs the dimensionless factor, $f_0$, introduced by
\citet{Duffell2015}). Note that the scaling of this relation is
  slightly different from the pebble isolation mass in
  \citet{2014A&A...572A..35L}, which depends on $h/r$, but not
  $\alpha$. However, a more recent prescription for the pebble
  isolation mass in \citet{2018arXiv180102341B} also includes a
  dependence on $\alpha$. A major difference of our results from these
  other studies is that our planet gap opening mass measurements are
  done specifically at a region in the disk where there is a
  transition in the value of $\alpha$ that causes a strong change in
  $\Sigma$. On the other hand, the results of
  \citet{2014A&A...572A..35L} and \citet{2018arXiv180102341B} are for
  disks with constant $\alpha$ and thus a relatively slowly varying
  $\Sigma_g$ profile.

To test equation~(\ref{eq:mg1}) and constrain $\phi_{\rm G,D}$, we now
extend the simulations of Paper III to investigate the gap opening
mass for different values of $\alpha$ and $\dot{m}$.  We perform a
series of FARGO \citep{2000A&AS..141..165M} hydrodynamic simulations on disks with an $\alpha$
viscosity transition. The inner region is still assumed to be an
MRI-active zone with $\alpha=0.01$. Beyond this is the dead zone with
a much lower value of $\alpha$. We implement the transition following
the same methods as Paper III, with the transition set to be at
0.1~AU, which is the location of the local pressure maximum before the
introduction of a planet in the disk. Then planets of different masses
are inserted in the disk and held fixed at this location (note Paper
III showed that growing planets would be trapped here given the
torques exerted by the disk). We study the cylindrically averaged
pressure profiles and identify the critical gap-opening mass for IOPF
as corresponding to the point when the local pressure maximum in the
disk becomes significantly displaced from the planet, i.e., by more
than its Hill radius,
\begin{eqnarray}
R_H & \equiv & \left(\frac{M_{p}}{3m_*}\right)^{1/3}r\\
 & \rightarrow & 1.00\times 10^{-3} M_{p,\oplus}^{1/3}m_{*,1}^{-1/3} r_{\rm 0.1AU}\:{\rm AU},\nonumber
\end{eqnarray}
where $M_{p,\oplus}\equiv M_{p}/M_\oplus$ is the planet mass
normalized to one Earth mass.

\begin{figure}
\centering
\includegraphics[width=0.5\textwidth]{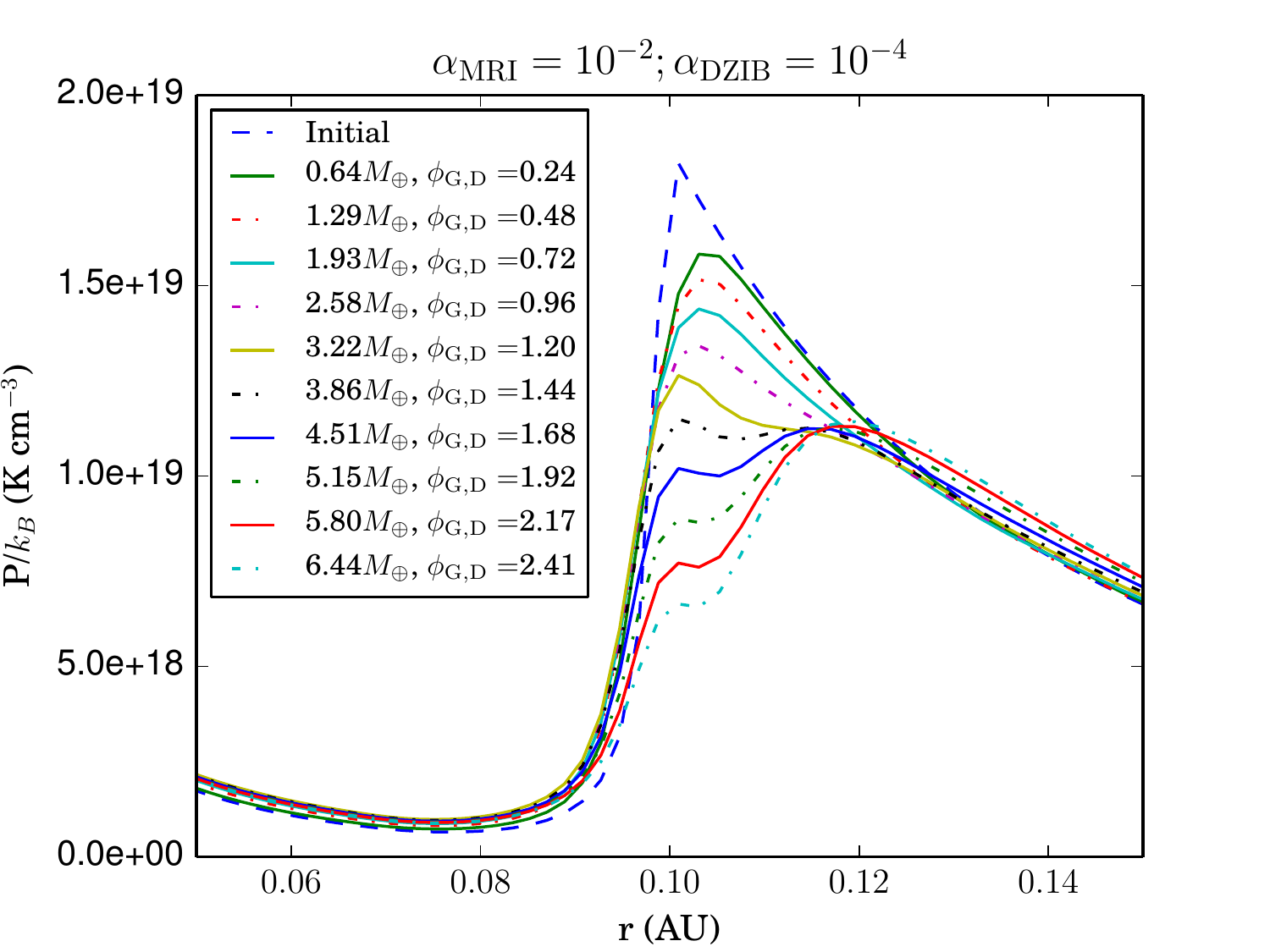}
\caption{
Steady state, cylindrically averaged disk midplane pressure profiles,
starting with an initial case with no planet. The pressure maximum,
due to the dead zone inner boundary, has been set at 0.1~AU. We set
$\alpha=0.01$ in the inner MRI-active disk, transitioning to $10^{-4}$
in the dead zone. Then the effects of adding planets of increasing
mass and held fixed at 0.1~AU are studied. These planets open gaps of
increasing depth, which eventually leads to displacement of the local
pressure maximum away from the planet. This occurs for $M_p =
3.86\:M_\oplus$, i.e., $\phi_{\rm G,D}=1.44$.
}
\label{fig:gap_1d}
\centering
\end{figure}

Figure \ref{fig:gap_1d} shows examples of these pressure profiles for
the case of a disk with $\dot{m} = 10^{-9}\:M_\odot\:{\rm yr}^{-1}$
and a transition from $\alpha=10^{-2}$ to $\alpha=10^{-4}$ in the
vicinity of 0.1~AU. We see in this case that a planet with mass of
$3.86\:M_\oplus$ is able to open a deep enough gap that the local
pressure maximum retreats significantly, i.e., by $\sim0.015\:$AU
(in this case, $\sim10$ Hill radii) from the planet.
 This mass scale corresponds to
a value of $\phi_{\rm G,D}=1.44$ in the normalization of the 
\citet{Duffell2015} gap opening mass (eq.~\ref{eq:mg1}). 
In the IOPF model this means that the next planet may form from a new
pebble ring at this location or somewhat further out if the MRI-active
region also spreads outwards in the lower densities enabled by gap
opening.

\begin{figure}
\centering
\includegraphics[width=0.48\textwidth]{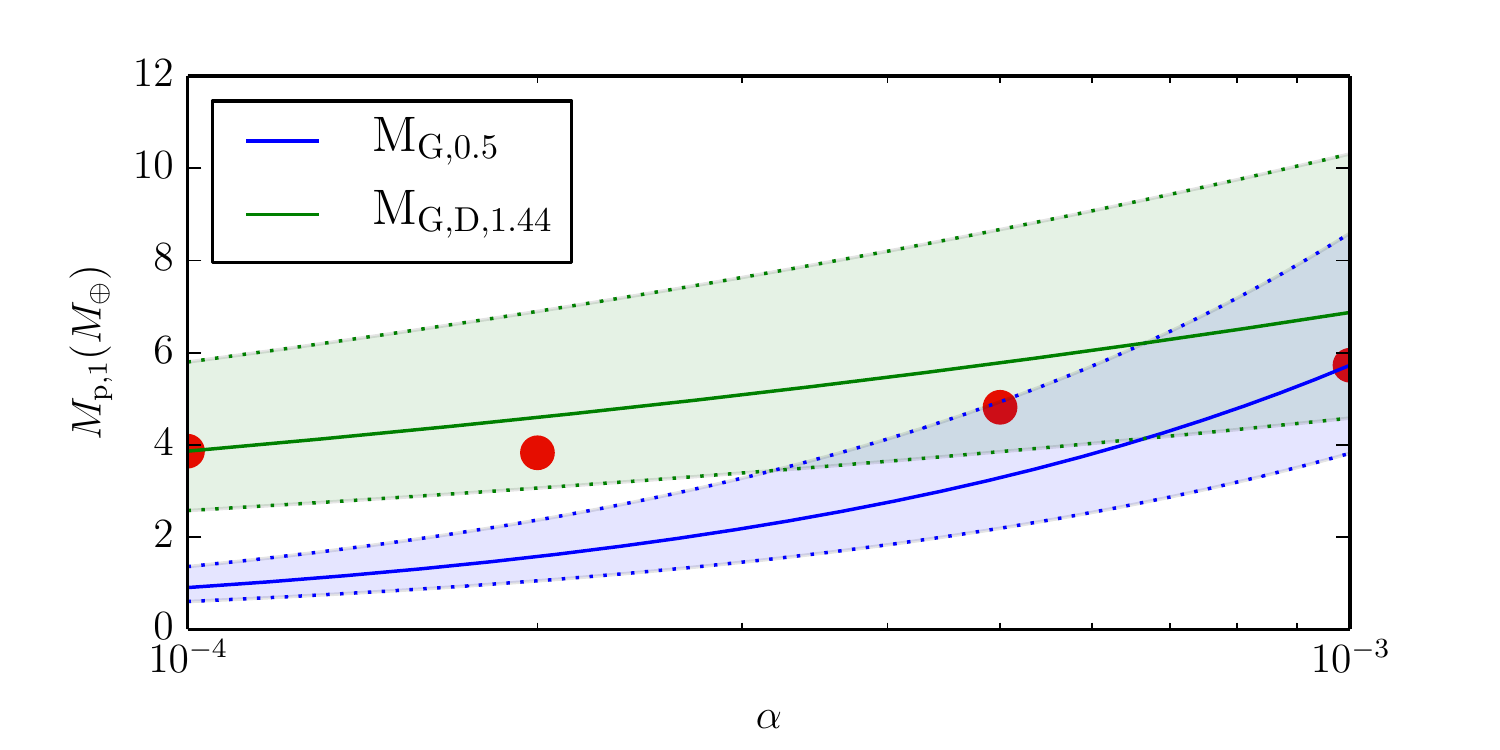}
\includegraphics[width=0.48\textwidth]{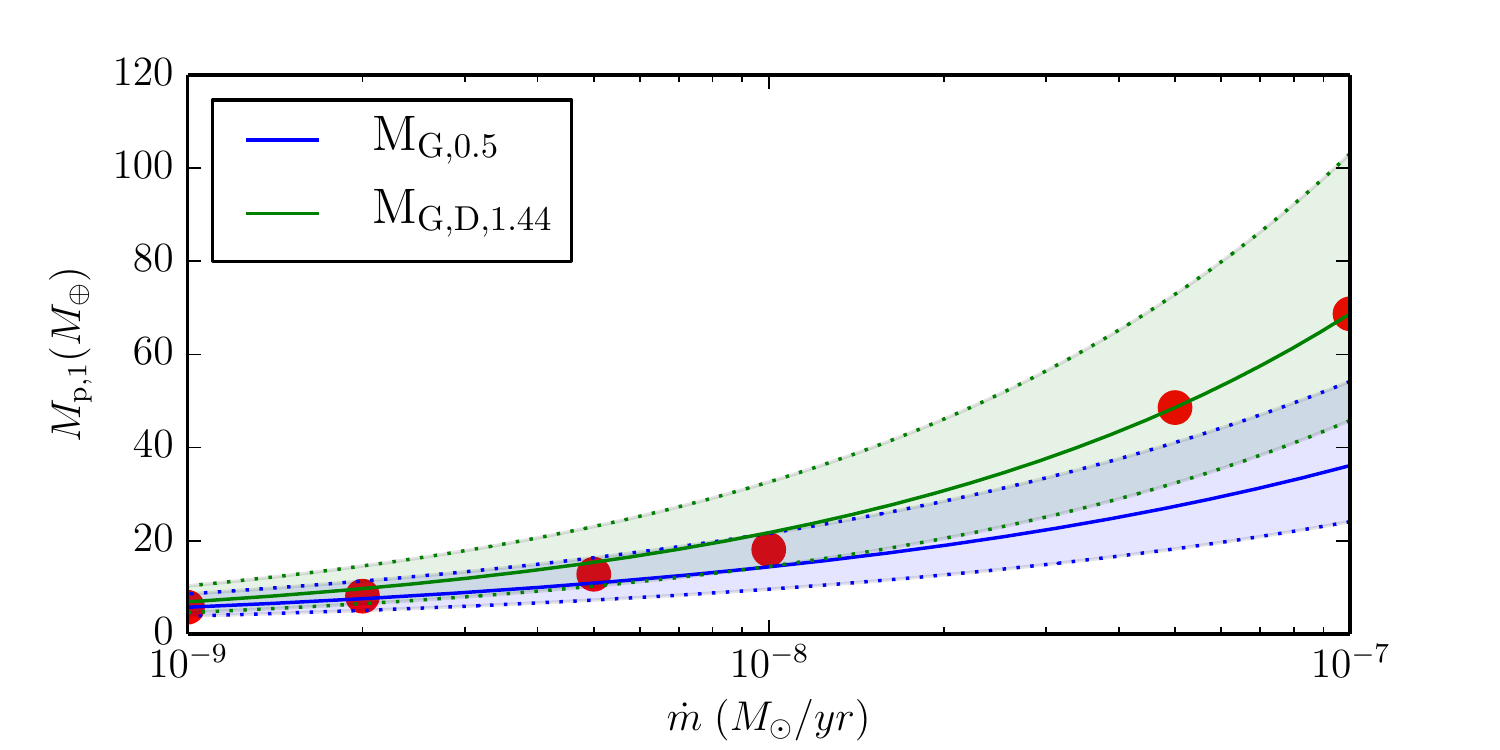}
\caption{
Planet gap opening mass estimates based on the viscous criterion
($M_{G,0.5}$, blue) and the \citet{Duffell2015} criterion ($M_{\rm G,D}$,
green) as a function of viscosity $\alpha$ (top panel) and accretion
rate $\dot{m}$ (bottom panel). The shaded area indicates a range of a
factor of 1.5 variation about the fiducial mass estimates. The red
dots are FARGO simulation results based on local pressure maximum
retreat. Note the \citet{Duffell2015} criterion is a better match with
the simulation results in the low $\alpha$ and high $\dot{m}$
regimes.}
\label{fig:gap_mass}
\centering
\end{figure}

Figure~\ref{fig:gap_mass}a shows the results of four different series
of simulations, including the example shown in Fig.~\ref{fig:gap_1d},
in which $\alpha$ in the dead zone is varied from $10^{-4}$ to
$10^{-3}$ and the gap opening masses (by the pressure maximum
displacement criterion) are assessed and compared to those predicted
by the simple viscous criterion (eq.~\ref{eq:MG}) and the
\citet{Duffell2015} estimate (eq.~\ref{eq:mg1}).
Figure~\ref{fig:gap_mass}b shows the results of a similar series of
gap opening simulations, but now varying $\dot{m}$ from $10^{-10}$ to
$10^{-7}\:M_\odot\:{\rm yr}^{-1}$. We find the simple viscous
criterion underestimates the planet mass required for displacing the
pressure maximum in both the low $\alpha$ and high $\dot{m}$
regimes. These are conditions when the disk is relatively denser and
hotter and its vertical scale height is relatively large compared to
the Hill sphere of the planet. Our results are consistent with those
of \citet{DufMac2013}: they discuss how gap profiles scale differently
when a planet's Hill radius is smaller than the disk scale
height. We note that our results are derived from 2D simulations,
  which should ideally be confirmed by 3D simulations that resolve the
  vertical structure of the disk to more accurately predict midplane
  pressure structures.

With our improved estimate of Vulcan planet mass set by gap opening
(eq.~\ref{eq:mg1}), following Paper II, we eliminate the accretion rate
term, $f_r\dot{m}_{-9}$, by combining it with
equation~\ref{eq:rinner}. Previously Paper II did this using the viscous
criterion gap opening mass to obtain $M_{p,1}=0.83
\phi_{G,0.5}\phi_{\rm DZIB,0.5}^{-9/10}\gamma_{1.4}\alpha_{-4}r_{\rm
  0.1AU}\:M_\oplus$. Now we find a revised mass versus orbital radius
relation for Vulcans:
\begin{eqnarray}
\label{eq:vulcan}
M_{p,1}&=&3.50\phi_{\rm G,D,1.44}\phi_{\rm DZIB,0.5}^{-9/8}\gamma_{1.4}^{5/4}\nonumber\\
 & \times &\alpha_{-4}^{1/2}m_{*,1}^{-1/4}r_{\rm 0.1 AU}^{5/4}\:M_\oplus.
\end{eqnarray}
We see that this has a slightly steeper scaling of $M_{p,1}\propto
r^{5/4}$ compared to the  Paper II result. The implications of this and
other revised IOPF predictions, such as the more general $M_p$ versus
$r$ relation of eq.~(\ref{eq:mg1}), for the comparison with observed
STIPs will be examined in a future paper.

\subsection{Formation Timescale}

Now that we have improved estimates for the masses of Vulcan planets,
we assess their formation timescales.
Consider a protoplanetary disk with mass accretion
rate of $\dot{m}$ and solids to gas mass ratio of $f_s$. We assume
$f_s=0.01$ as a fiducial value. The solids are divided into two parts:
(1) sub-millimeter or smaller sized dust grains
that are well coupled to the gas; (2) larger, typically centimeter
sized ``pebbles''
that can decouple from the gas flow and can be subject to substantial
radial drift due to gas drag. 

Pebbles are expected to arise from coagulation of dust grains, mainly
via Brownian motion and turbulent mixing. The maximum steady state rate of pebble
delivery to a particular radial location in the disk is if most solids
are in the form of pebbles, i.e., $f_p = \dot{m}_p/\dot{m}_s\simeq 1$.
The accretion limited pebble supply rate is then simply:
\begin{eqnarray}
\dot{m}_p &=&10^{-11} f_p f_{s,0.01} \dot{m}_{-9}\: M_\odot\:{\rm yr}^{-1}\\
&=& 3.33\times 10^{-6}f_p f_{s,0.01} \dot{m}_{-9}\: M_{\oplus}\:{\rm yr}^{-1}.\nonumber 
\end{eqnarray}

Thus we have an estimate of the planet formation time if limited by a
steady accretion rate of pebbles:
\begin{eqnarray}
t_{\rm form}&=&M_{\rm G,D,1.44}/\dot{m}_p \nonumber\\
&=&f_s^{-1}f_0^{-1/2}\frac{3^{3/4}}{2^{7/4}}\left(\frac{\mu}{\gamma k_B}\right)^{-1}\left(\frac{\kappa}{\sigma_{\rm SB}}\right)^{{1}/{4}} \alpha^{1/4}\nonumber\\
&\times&G^{-7/8}m_*^{1/8}f_r^{1/2}\dot{m}^{-1/2}r^{{1}/{8}},\\
&\rightarrow& 
1.16 \phi_{\rm G,D,1.44}f_{s,-2}^{-1} \nonumber\\
&\times& \gamma_{1.4}\kappa_{10}^{1/4}\alpha_{-4}^{1/4}m_{*,1}^{1/8}f_r^{1/2}\dot{m}_{-9}^{-1/2}{r_{\rm 0.1 AU}}^{1/8}\: \rm Myr \nonumber
\end{eqnarray}
For solar type stars, estimates of the lifetimes of protoplanetary
disks are $\sim1$ to 10 Myr, with median values of $\sim$3 Myr,
\citep[e.g.,][]{2011ARA&A..49...67W, 2015A&A...576A..52R}. Since there
are usually $\gtrsim3$ planets in a typical STIP, this $\sim$1~Myr
formation time for the first planet may give an interesting constraint
on IOPF models: e.g., to form $\sim3$ planets may take a timescale
that is very similar to typical disk lifetimes. Note, the above planet
formation timescale is not very sensitive to model parameters, such as
opacity ($\kappa$), viscosity ($\alpha$) and location ($r$).

\subsection{Pebble Supply Reservoir}

The radius of the ``pebble supply reservoir,'' $r_{\rm res,1}$, that is
needed to construct a Vulcan planet can now also be evaluated by
equating this to the disk radius that encloses a solid mass that is at
least as large as $M_{p,1}$. We define
\begin{eqnarray}
M_{p,1} = \epsilon_{p,1} M_d(<r_{\rm res,1})
\end{eqnarray}
and show some example estimates of $r_{\rm res,1}$ in
Table~\ref{tab:zone}, given our disk models (e.g., shown in
Fig.~\ref{fig:disk}). Here $\epsilon_{p,1}$ denotes planet formation
efficiency, i.e., the fraction of solids within the reservoir that
finally becomes part of planet. The size of the supply reservoir is
sensitive to the value of $\alpha$ adopted for the disk model. For the
fiducial case of $\alpha=10^{-4}$ (i.e., potentially appropriate for
dead zone conditions) and with $\dot{m}=10^{-9}\:M_\odot\:{\rm
  yr}^{-1}$, we find $r_{\rm res,1}\simeq 3\:$AU. However, if the
value of $\alpha$ is larger, then $r_{\rm res,1}$ grows, potentially
to several tens of AU, both because of the larger planet mass and the
lower mass surface density of the disk.

\section{Single pebble evolution model}
\label{S:single}

Here we describe a model of single pebble evolution in the disk, i.e.,
involving radial drift due to gas drag and growth of the pebble by
sweeping up small dust grains. This model is the basis of that used in
the next section to predict the global evolution of the pebble
population. Here we will first use the single pebble evolution model
for simple estimates of the supply timescales needed to form
Vulcan planets.

\begin{figure}[t]
\centering
\includegraphics[width=0.48\textwidth]{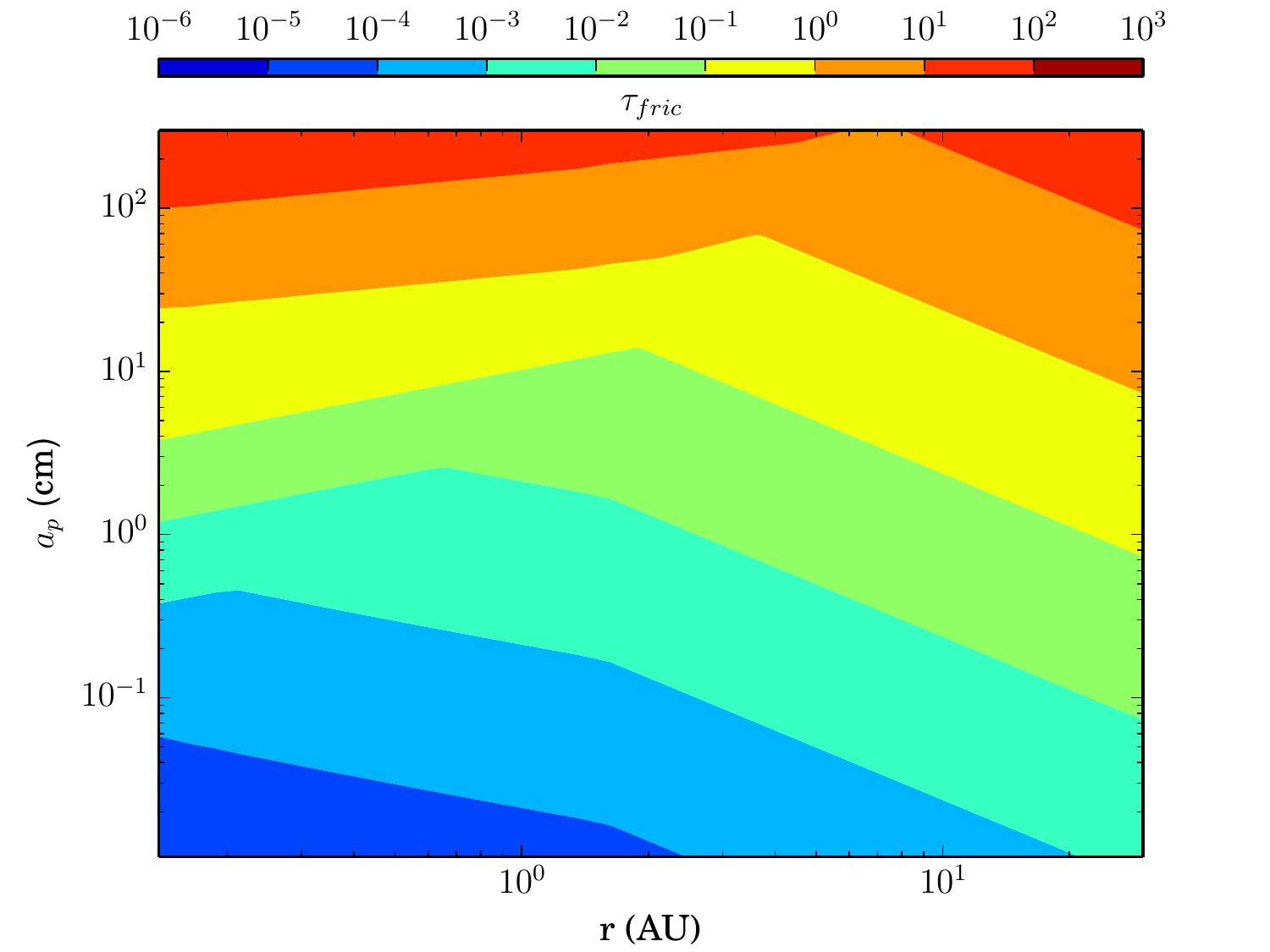}
\caption{
Dimensionless friction timescale, $\tau_{\rm fric}$, map of pebbles in
our fiducial disk model ($\dot{m}=10^{-9}\:M_\odot\:{\rm yr}^{-1}$,
$\alpha=10^{-4}$).  
}
\label{fig:tau}
\centering
\end{figure}

The pebble evolution model is based on that presented by
\citet{2014IAUS..310...66H}. It includes four different drag regimes
(Epstein; and three Stokes regimes, depending on the relative radius
of the pebble, $a_p$ to the mean free path for collisions in the gas,
$\lambda$, and the Reynolds number, $Re$) to evaluate the drag force
$F_D$. We then calculate the gas drag frictional timescale of a pebble
of mass $m_p$ moving at speed $v_p$ relative to gas as $t_{\rm
  fric}=(m_pv_{p})/F_{D}$:
\begin{equation}
t_{\rm fric}=
\begin{cases}
\rho_p a_p / (\rho v_p) & \text{if }  a_p < 9\lambda/4 \\
2\rho_p a_p^2 / (9\nu\rho) & \text{if } a_p > 9\lambda/4 \text{ \& } {Re} <1\\
(\rho_p a_p / [9\rho v_{p}]) (2av_{p}/\nu) ^{0.6} & \text{if } 1<{Re}<800\\
8\rho_p a_p / (1.32\rho v_{p}) & \text{if } {Re} > 800.
\end{cases}\label{eq:drag}
\end{equation}
Then the radial drift of pebble relative to gas is \citep{2007astro.ph..1485A}:
\begin{equation}
v_{r,p}  \simeq  -k_P(c_s/v_K)^2 (\tau_{\rm fric}+\tau_{\rm fric}^{-1})^{-1} v_K,
\end{equation}
where $k_P$ is power-law index of pressure radius
relation in $P=P_0(r/r_0)^{-k_P}$, $\tau_{\rm fric}\equiv \Omega_K t_{\rm fric}$ is the
dimensionless friction timescale and $\Omega_K$ is the orbital angular
frequency. An example of the values of $\tau_{\rm fric}$ as a function
of pebble size and disk radius for our fiducial disk model
($\dot{m}=10^{-9}\:M_\odot\:{\rm yr}^{-1}$, $\alpha=10^{-4}$) is shown
in Figure~\ref{fig:tau}. We can see from this figure that efficient
radial drift, i.e., when $\tau_{\rm fric}\sim 1$, occurs for
$\sim$cm-sized pebbles in the outer disk region, which will be
important for the models we consider below that are based on steady
injection of pebbles at this outer disk scale (see \S\ref{S:model}).

\begin{deluxetable*}{cccccc}[h]
\tablecaption{Pebble Supply Reservoir Radii and Drift Timescales for Vulcan Planets\label{tab:zone}}
\tablehead{disk model & $M_{p,1}$ & $r_{\rm 1200K}$ & $r_{\rm res,1}$ (AU) & $t_{\rm drift}$ (no growth)  & $t_{\rm drift}$ (w/ sweep-up growth)\\
 ($\dot{m}_{-9}, \alpha$) & ($M_\oplus$) & (AU) & ($\epsilon_{p,1}=1,0.5,0.25$) & ($10^4\:$yr, $\epsilon_{p,1}=1,0.5,0.25$) & ($10^4\:$yr, $\epsilon_{p,1}=1,0.5,0.25$) }
\startdata
$0.1, 10^{-4}$ & 0.96 & 0.060 & 3.35, 6.87, 14.3 & 6.20, 7.23, 8.18 & 0.748, 1.17, 1.71\\
$0.1, 10^{-3}$ & 1.59 & 0.036 & 19.1, 40.2, 84.5 & 0.994, 1.22, 1.94 & 0.474, 0.697, 1.41\\
$1, 10^{-4}$ & 3.43 & 0.167 & 1.75, 2.95, 5.45 & 16.3, 24.6, 33.6 & 0.798, 1.35, 2.35\\
$1, 10^{-3}$ & 5.73 & 0.100 & 6.67, 13.7, 28.5 & 5.43, 6.37, 7.23 & 1.14, 1.67, 2.28\\
$10, 10^{-4}$ & 12.3 & 0.464 & 1.93, 2.90, 4.47 & 17.7, 29.9, 50.4 & 0.671, 1.14, 2.08\\
$10, 10^{-3}$ & 20.6 & 0.278 & 3.73, 6.14, 11.1 & 12.6, 19.8, 27.7 & 1.39, 2.40, 3.98
\enddata
\tablecomments{
$t_{\rm drift}$ is evaluated for pebbles of initial radius of 1~mm,
  with the trajectory followed from $r_{\rm res,1}$ to the DZIB where
  $r=r_{1200\rm K}$ (for $\phi_{\rm DZIB}=0.76$).}
\end{deluxetable*}

The pebble evolution model of \citet{2014IAUS..310...66H} involves
``sweep-up growth'' of pebbles by accretion of small dust grains (we
add pebble-pebble coagulation in the next section). Similar models
have been proposed by \citet{Safranov1972} and also been discussed by
\citet{DD2005} in the context of dust growth during vertical
settling. \citet{Windmark2012a} considered such a model as a mechanism
by which cm-sized ``seeds'' can overcome the bouncing/fragmentation
barrier to form planetesimals by sweeping up smaller grains. 

Like \citet{2014IAUS..310...66H}, we assume a pebble sweeps up all the
dust within its geometric cross section if it is in the Epstein drag
regime, where pebble size is comparable to a gas molecule's mean free
path. However, when a pebble enters the Stokes regime, the gas behaves
more like a continuous fluid, forming a pressure wake in front of the
pebble. We thus assume sweep up growth stops, as small dust grains are
deflected by this pressure wake and flow away following gas
streamlines.

To implement this model, in each time step $\delta t$, the mass growth of a
pebble is:
\begin{eqnarray}
\delta m_p&=&\pi a_p^2 v_{\rm rel} \rho_d \delta t.
\label{eq:sweep_up_mass} 
\end{eqnarray} 
Here $\rho_d$ is mass of dust per unit volume in the disk, i.e., 1\%
of the gas density, $\rho$, for our fiducial initial conditions.
$v_{\rm rel}$ is the relative velocity between a pebble and its
surrounding gas, which is given by:
\begin{eqnarray}
v_{\rm rel}=\left(1+\tau_{\rm fric}^2/4\right)^{1/2}v_{\rm r,p}.
\end{eqnarray} 

\begin{figure*}
\centering
\includegraphics[width=0.9\textwidth]{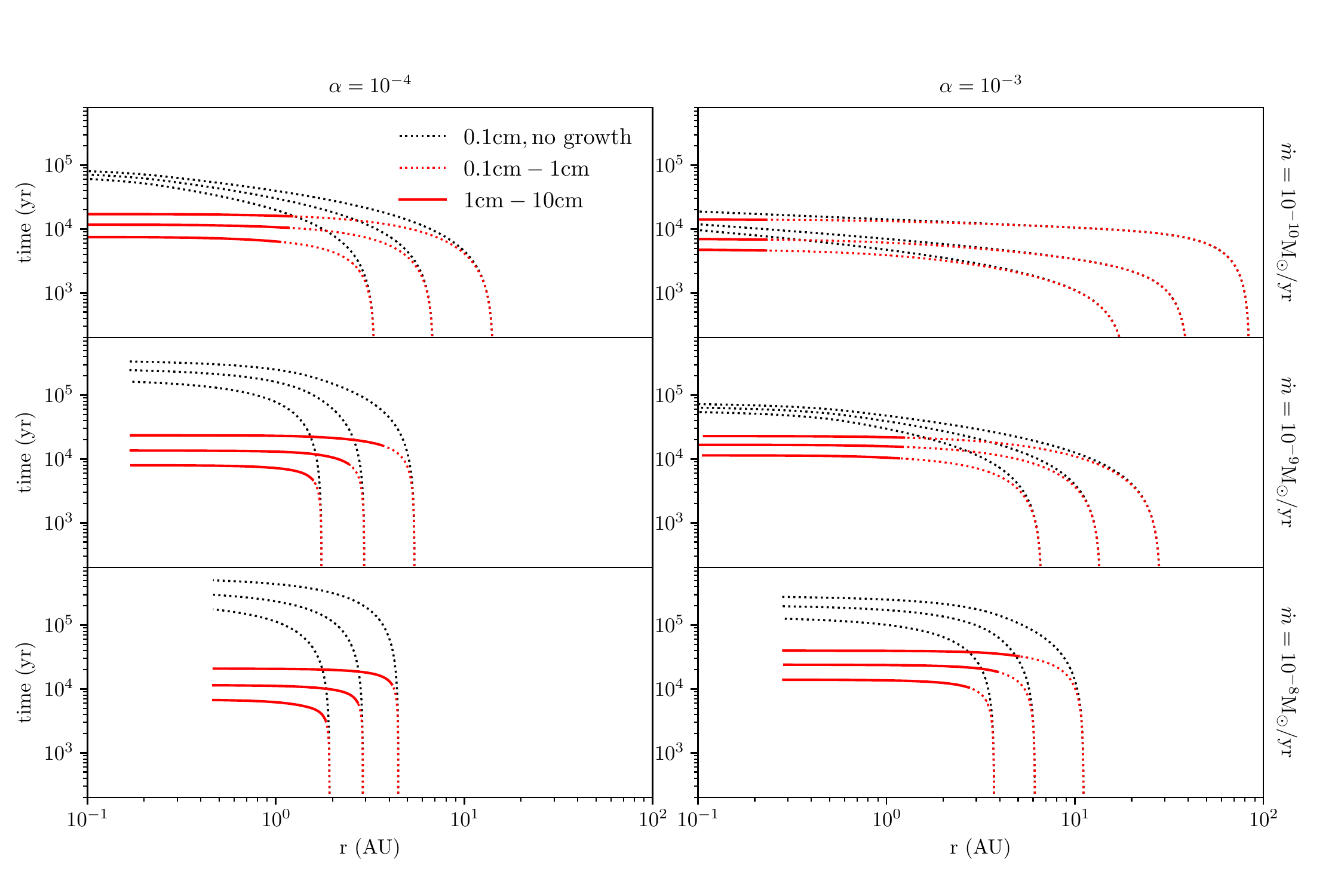}
\caption{
Radial trajectories in time of pebbles with initial radii of 0.1~cm,
starting from various supply reservoir outer boundaries for Vulcan
planets with $\epsilon_p=1, 0.5, 0.25$ (see Table~\ref{tab:zone}). The
black dotted lines show models with no growth of the pebbles,
i.e., they keep a constant radius of 0.1~cm. The red lines show
models with growth via Stokes-limited sweep-up of small dust
grains.  Different styles of red lines represent different ranges
  of pebble sizes during this growth: dashed lines cover the size
  range of 0.1 to 1cm; solid lines cover 1 to 10 cm.  Left column
shows disks with $\alpha=10^{-4}$; right column with
$\alpha=10^{-3}$. Top, middle and bottom rows show models with
$\dot{m}=10^{-10}, 10^{-9}, 10^{-8}\:M_\odot\:{\rm yr}^{-1}$,
respectively. Pebble trajectories are followed until an inner radius
is reached where $r=r_{1200\rm K}$, $\phi_{\rm DZIB}=0.76$, expected
to be the location of the DZIB.}
\label{fig:single_peb}
\centering
\end{figure*}

We now calculate example trajectories of pebbles of initial radius of
1~mm that start at a radius in the disk equal to that of the pebble
supply reservoir for a Vulcan planet, $r_{\rm res,1}$ (see
Figure~\ref{fig:single_peb}). The trajectories are shown for the cases
with and without sweep-up growth, i.e., the latter being for pebbles
of constant radius of 1~mm. These calculations are done for six
different disk models, i.e., $\dot{m}_{-9}=0.1, 1, 10$ and
$\alpha=10^{-4}, 10^{-3}$. In each disk, three different starting
radii for $\epsilon_{p,1}=1,0.5,0.25$ are considered. These properties
and the total drift times, $t_{\rm drift}$, for all these cases are
listed in Table~\ref{tab:zone}.

Figure~\ref{fig:single_peb} and Table~\ref{tab:zone} illustrate
several points. First, note the dependence of Vulcan planet masses,
$M_{p,1}$, and the pebble supply reservoir outer radii, $r_{\rm
  res,1}$, on disk properties. These planet masses range from about
$1\:M_\oplus$ in the case of a low accretion rate, low viscosity disk,
to about $20\:M_\oplus$ in a disk with a $100\times$ higher accretion
rate and $10\times$ higher viscosity. Since disk mass surface
densities are lower for higher viscosity disks, the corresponding
reservoir radii also become larger in such cases. Thus $r_{\rm res,1}$
varies from just 1.75~AU (actually occurring in the fiducial case if
$\epsilon_{p,1}=1$) to almost 100~AU in the low $\dot{m}$, high
$\alpha$, low $\epsilon_{p,1}$ model.

Then the drift times from these radii to the location of planet
formation, i.e., $r_{\rm 1200K}$, depend on whether the pebble is
allowed to grow by sweep-up growth of small dust grains. Without
growth, in the $\alpha=10^{-4}$ disks, we see that $t_{\rm
  drift}\gtrsim10^5\:$yr. It takes longer to drift inwards in the high
$\dot{m}$ disks, even though starting closer in, because the pebbles
are in different drag regimes leading to different drag frictional
timescales (eq.~\ref{eq:drag}). This pattern is mirrored in the higher
$\alpha$ disks, where we see that the no growth drift times can be as
short as $\sim10^4\:$yr even from $\sim$100~AU. Allowing pebbles to
grow results in shorter drift timescales, since the pebbles approach
sizes that lead to maximal drag force, i.e., minimal frictional
timescales, leading to efficient radial migration, which then further
enhances pebble growth. We notice that in all the six disk models
considered the values of $t_{\rm drift}$ with sweep-up growth are
quite similar, i.e., $\sim 10^4\:$yr. The pebbles can grow to sizes
$\sim 1$ to 10~cm.

Thus we conclude that for a wide range of disk parameters ($\dot{m}\in
[10^{-10},10^{-8}]\:M_\odot\:{\rm yr}^{-1}$, $\alpha \in
[10^{-4},10^{-3}]$), the radial drift time scale is one or two orders
of magnitude shorter than $t_{\rm form}$, based on an estimate of a
steady-state supply rate of pebbles. This reflects the extreme limit
of the pebble supply rate being boosted above the accretion limited
rate because of the net radial drift of pebbles with respect to
gas. It also assumes that the pebbles starting from within $r_{\rm
  res,1}$ will be able to sweep up a large fraction of the total solid
mass, i.e., of dust, in this inner disk region. To advance beyond
these simple estimates, we need to construct a global model of pebble
evolution and supply in the disk.

\section{Global Pebble Evolution Model}
\label{S:model}

\subsection{Numerical Implementation}

Here we describe the basic algorithm used in an Eulerian pebble
evolution model, including radial drift and growth (sweep-up plus
coagulation). We sample a discrete grid in radial distance ($r$) and
pebble radius ($a_p$). At any given radius, the model is intended to
approximate conditions applicable to 
pebbles drifting inwards near the disk midplane.

We divide solids in the disk into two groups: larger particles treated
as pebbles; smaller particles treated as dust. Dust represents
particles smaller than a certain size threshold, for which we adopt
$a_{p,{\rm min}}=0.01$~cm (as a fiducial choice), while pebbles are
particles divided in different size bins.  During radial drift,
pebbles can grow by sweeping up dust (i.e., the Stokes-limited
sweep-up growth model described in the previous section) and by
coagulation with other pebbles.
    
The basic numerical approach in an ``Eulerian'' model is to find the
fraction of material that is transported between neighboring cells
within each time step. Considering pebble evolution, this fraction at
radial grid cell $i$ and pebble size grid cell $j$, $f_{p}(i,j)$ is
calculated as the sum of the radial drift fraction in $r$ space,
$f_{p,r}(i,j)$ and size ($a_p$) space $f_{p,a}(i,j)$.
The radial drift fraction of mass moved to the next radial grid, i.e.,
the mass fraction of pebble size $j$ at radial grid $i$,
$f_{p,r}(i,j)$ is the ratio between radial drift distance to radial
width of the ring shaped grid:
\begin{eqnarray}
f_{p,r}(i,j)&=& v_{r, \rm i,j} \Delta t / \Delta r_{\rm ring},
\end{eqnarray}
where $\Delta r_{\rm ring}$ is the width of each radial grid, $\Delta
t$ is time step, and $v_{r, \rm i,j}$ is radial drift velocity of
pebble size $j$ at the center of the grid. This ``center of box''
first order approximation works because the variation of disk
properties within each grid is very minor, especially with a
logarithmic radial grid set up.

The size evolution of pebbles is more complicated. The difference
between largest and smallest pebbles within one size bin becomes
larger due to the nature of sweep up growth. At size bin $j$, the
minimum size is denoted as $a_{\rm p,i,jmin}$, and the maximum is
$a_{\rm p,i,jmax}$. Within one time step, a pebble with size $a_{\rm
  p,i,jmin}$ can grow to size $a_{\rm p,i,jmin}^\prime$, by sweeping
up dust of mass $\delta m$ from eq.~\ref{eq:sweep_up_mass}:
\begin{equation}
a_{\rm p,i,jmin}^\prime=\left(a_{\rm p,i,jmin}^3+\frac{3\delta
  m}{4\pi}\right)^{1/3}.
\end{equation}
Similarly, a $a_{\rm p,i,jmax}$ sized pebble can grow to $a_{\rm
  p,i,jmax}^\prime$.  Between size $a_{\rm p,i,j}$ and $a_{\rm
  p,i,j+1}$, there exists a size $a_{\rm p,i,jmid}$, that can grow to
$a_{\rm p,i,jmax}$ in this time step, $\Delta t$. The value of $a_{\rm
  p,i,jmid}$ is obtained from linear interpolation of $a_{\rm
  p,i,jmin}$ and $a_{\rm p,i,jmax}$. Thus, pebbles below size $a_{\rm
  p,i,jmid}$ will stay in current size bin, while pebbles above
$a_{\rm p,i,jmid}$ move to next size bin. With the assumption that
pebble mass is distributed evenly within each size bin, we obtain the
fractional mass in size space that grows to the next size bin:
\begin{eqnarray}
f_{p,a}(i,j)&=& \frac{\left(a_{\rm p,i,jmax}-a_{\rm p,i,jmid}\right)}{\left(a_{\rm p,i,jmax}-a_{\rm p,i,jmin}\right)}.
\end{eqnarray}

To make the code more efficient, at each mesh in $r$ space, the time
step is self adjustable, depending on the drift rate and growth rate
of pebbles. The whole disk shares a larger time step for
synchronization, while each mesh has its own sub time step that is a
simple fraction ($1/n$) of the larger step. To achieve this, virtual
rings are implemented between each mesh, recording fluxes from the
outer mesh and supplying to inner meshes in each sub step.

\begin{figure}[h]
\centering
\includegraphics[width=0.45\textwidth]{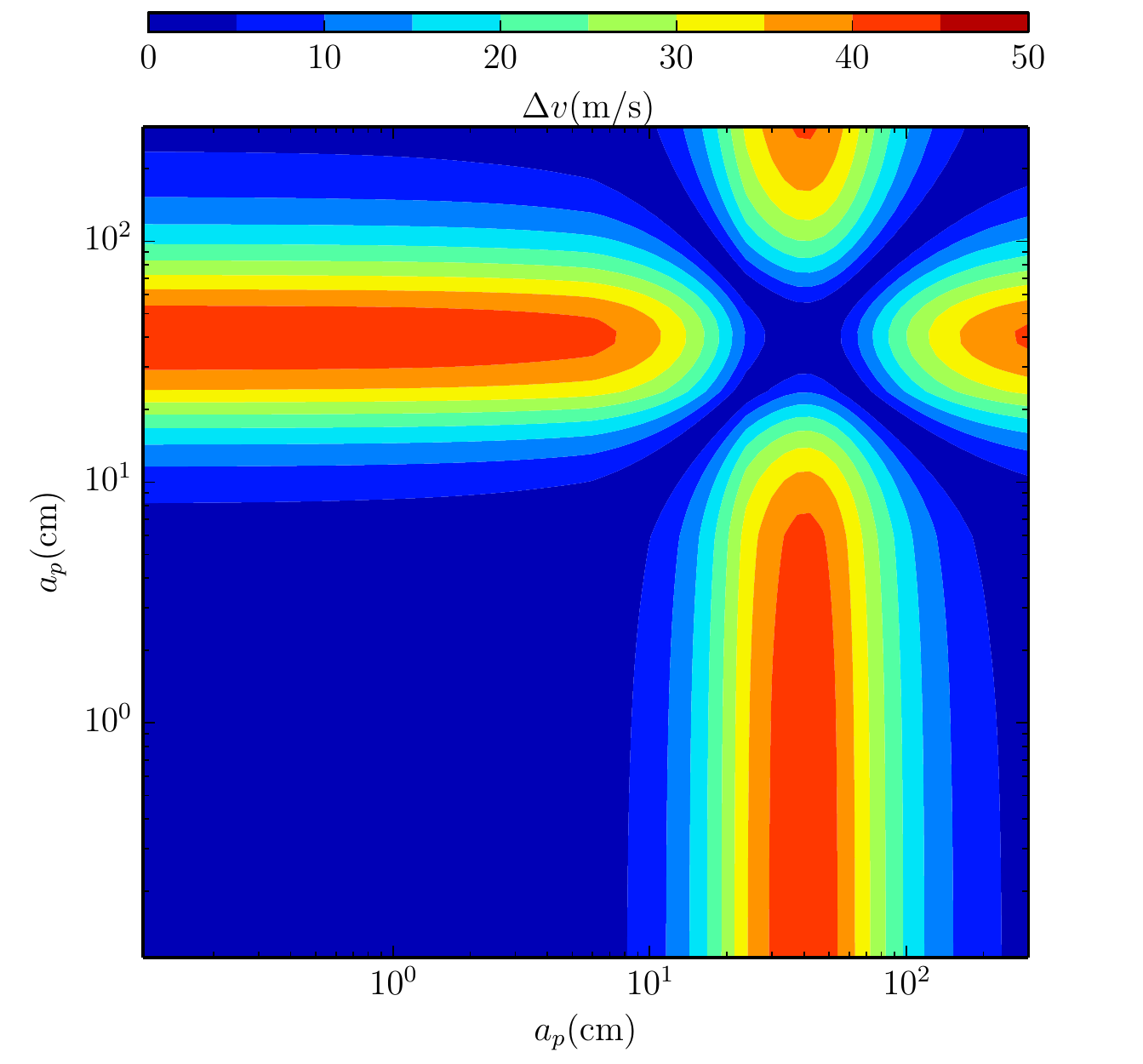}
\caption{
Example of total disk plane relative velocities between different
pebbles due to radial and cylindrical drift in the fiducial disk model
($\dot{m}_{-9}=1$, $\alpha=10^{-4}$) at 1~AU. These relative
velocities are used in the pebble-pebble coagulation model.
}
\label{fig:delta_v}
\centering
\end{figure}

A simple pebble-pebble interaction algorithm is also implemented. Here
we ignore effects like Brownian motion, disk turbulence or vertical
settling, so that only radial and cylindrical drift are considered for
setting interaction velocities.
When pebble size group $j$ meets group $k$, the number of two body
interactions leading to coagulation, $N_c$, per unit area per time
step, $\Delta t$, is calculated as \citep[see][]{Birnstiel2010}:
\begin{eqnarray}
N_c=S_{j,k}\Delta v_{j,k} \Delta t p_c N_j N_k /\sqrt{2\pi(h^2_j+h^2_k)},
\end{eqnarray}
where $S_{j,k}=\pi(a_{p,j}+a_{p,k})^2$ is the coagulation cross section, 
$\Delta v_{j,k}$ is the relative velocity between pebble
group {\it j} and group {\it k}, $N_j$ and $N_k$ are surface number
densities of pebbles in these groups, $h_j$ and $h_k$ are pebble
vertical scale heights, calculated as $h_j=h/\sqrt{1+\tau_{\rm
    fric,j}/\alpha}$, and $p_c$ is the coagulation
efficiency given by:
\begin{equation}
p_c=
\begin{cases}
1, & \text{if}\ \Delta v \leq 10 {\rm m/s} \\
3 - (\Delta v / 5 \:\rm m/s), & \text{if} \ 10 {\rm m/s} < \Delta v \leq 15 {\rm m/s}\\
0,&  \text{if}\ \Delta v \geq 15 {\rm m/s}.
\end{cases}
\end{equation}
Thus when the relative speed is $>15$~m/s, there is a ``fragmentation
barrier'' that prevents further pebble growth. 

Figure~\ref{fig:delta_v} shows an example of the relative velocities
occurring between different size groups at an 1~AU location in the
fiducial disk with $\dot{m}=10^{-9}\:M_\odot\:{\rm yr}^{-1}$ and
$\alpha=10^{-4}$. These range up to $\sim$40~m/s for pebbles with
$a_p\sim40\:$cm, but, for typical sizes of $\sim1$ to 10~cm, are
$\lesssim10\:$m/s.

The final step of modeling coagulation is that of redistributing the
masses of the coagulation products. Each pebble group has a certain
range of sizes (i.e., masses), and the width of this range would be
enlarged by coagulation, as the product's minimum size is set by the
combination of the smallest pebbles from both groups and the maximum
size by combination of the largest pebbles. So the coagulated mass
will be distributed into size bins ranging from $(a_{\rm
  p,i,jmin}^3+a_{\rm p,i,kmin}^3)^{1/3}$ to $(a_{\rm
  p,i,jmax}^3+a_{\rm p,i,kmax}^3)^{1/3}$.

The coagulation time step also follows a sub-step algorithm, in each
mesh, with the condition that $\Delta t$ is chosen so that the
fraction of pebbles (by number or mass) that coagulate is no larger
than 0.02. 

For our simulation domain, we only consider pebble evolution in the
region outward of the pressure maximum of the DZIB, i.e., from about
0.1 AU to 30 AU.

The outer boundary of 30 AU is chosen as a representative outer disk
scale, which contains a large enough dust reservoir to form a system
of planets for the various disk models considered. However, this
choice is somewhat arbitrary and, as we will see, we will mainly focus
on the results of models that involve a steady injection of small
pebbles at this outer boundary, so the initial reservoir of solid
material is not particularly important.

The radial spatial resolutions of the model is set by
160 logarithmic divided grids, from 0.1 AU to 30 AU.
 We follow pebbles from a minimum particle radius of
$a_{p,{\rm min}}$, while solids below this limit are considered to be
``dust.'' While $a_{p,{\rm min}}=0.01$~cm is the fiducial choice, we
also consider models with $a_{p,{\rm min}}=0.1$~cm to explore the
dependence of our results on this choice. Note, we use logarithmically
spaced grids in pebble size space: $a_{p,j}=a_{p, {\rm
    min}}\times10^{j\times0.05}$.  At the start of simulation, the
dust mass surface density $\Sigma_d$ is set to be 1\% of the gas mass
surface density, $\Sigma_g$. The initial pebbles are given a
mass surface density of $\Sigma_p$ that is 1\% of $\Sigma_d$ (in the
fiducial case; variation of this parameter is also explored). The
distribution of the radii of the initial/injected pebbles are set to
range from $a_{p,{\rm min}}$ to 0.3~cm with a number density versus
size
distribution following the equilibrium distribution of
\citet{Birnstiel2015}:
\begin{equation}
{\rm d} n_p(a_p)  \propto a_p^{-5/8} {\rm d} a_p.
\label{eq:sdf}
\end{equation}
At the outer boundary, the disk is supplied with dust and pebbles
given the steady state accretion rate and our adopted value of
$f_s=0.01$,  
and the fiducial pebble to dust mass ratio is again 1\% for the
injected solids.

\subsection{Fiducial Model Results \& Effects of Accretion Rate, Initial Pebble Distribution, \& Minimum Pebble Size}

\begin{figure*}[t]
\centering
\includegraphics[width=1.0\textwidth]{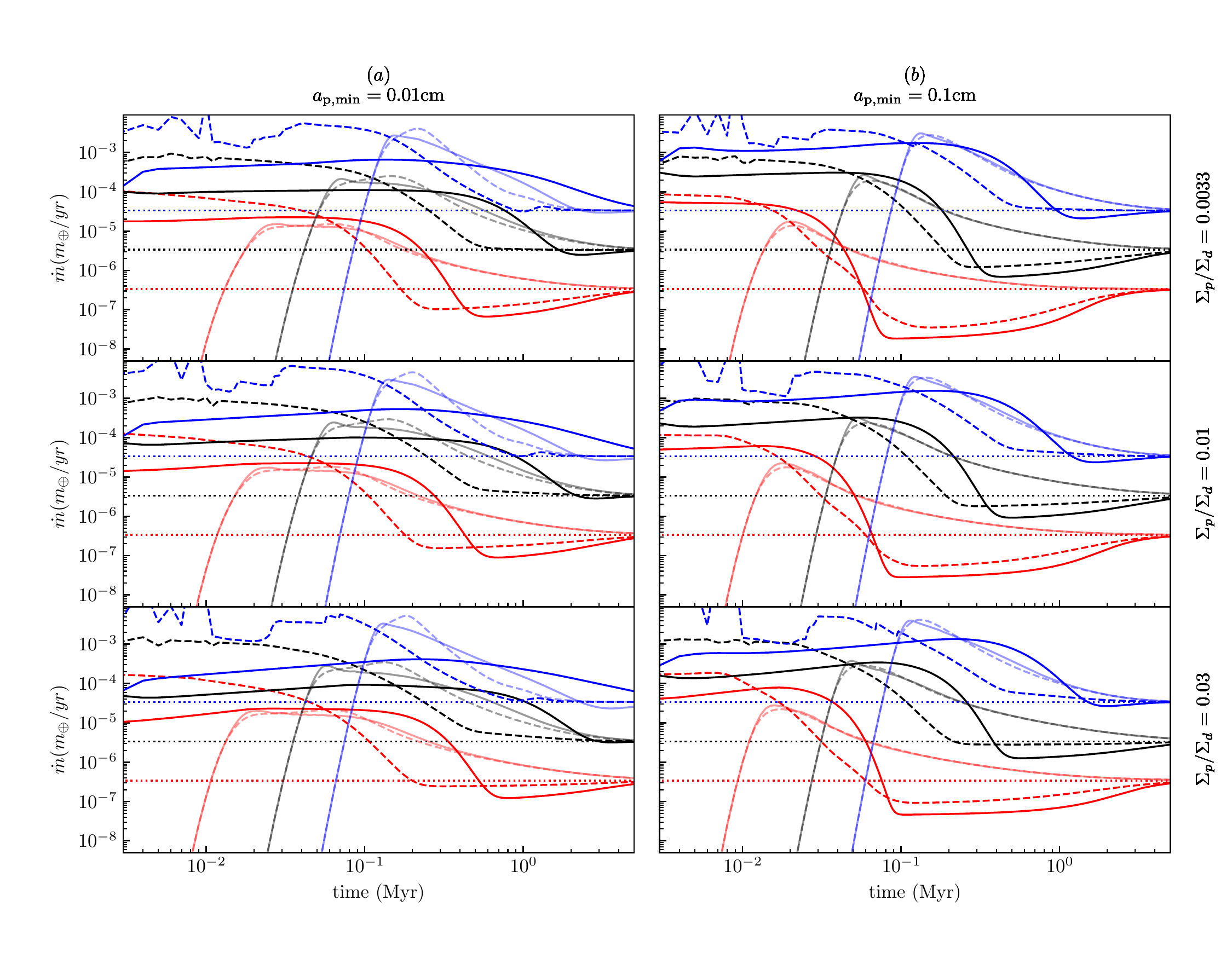}\\
\caption{
{\it (a) Left:} Pebble accretion rate to the inner disk ($r=0.1\:$AU)
region versus time for models with minimum pebble radius $a_{p,{\rm
    min}}=0.01\:$cm. The {\it upper, middle} and {\it lower} panels
show different initial pebble to dust ratios
$\Sigma_p/\Sigma_d=0.0033,0.01,0.03$, respectively, and the same
values are also used for the injected pebble to dust ratios. Note
  that all models assume a constant solids (dust $+$ pebbles) to gas
  ratio of 0.01.
In each panel, 3 different accretion rates are color coded.
The solid black line shows the model of pebble
evolution with only sweep-up growth for the case of the
$\dot{m}=10^{-9}\:M_\odot\:{\rm yr}^{-1}$ disk with
$\alpha=10^{-4}$. The dashed black line shows the equivalent result
for the model that also includes pebble-pebble coagulation. The
horizontal black dotted line shows the steady state pebble accretion
rate if all solids are in the form of pebbles. Equivalent results for
higher ($\dot{m}=10^{-8}\:M_\odot\:{\rm yr}^{-1}$, blue) and lower
$(\dot{m}=10^{-10}\:M_\odot\:{\rm yr}^{-1}$, red) accretion rate disks
are shown. The lines in lighter color shading show results of disks
that are initially empty of pebbles, i.e., pebbles only appear via
injection at the outer boundary at 30~AU.
Note the pebble accretion rates in the simulated disks can be larger
than the steady state rates because of the sweep-up and delivery of
the initial dust reservoir in the disk, but the asymptotic behavior at
late times is towards this steady state rate.
{\it (b) Right :} As (a) but now for $a_{p,{\rm min}}=0.1\:$cm.}
\label{fig:peb_flux1e-4}
\centering
\end{figure*}

We first present results in Figure~\ref{fig:peb_flux1e-4} for the time
evolution of the mass flux of pebbles delivered to the inner disk
inner boundary ($r=0.1\:{\rm AU}$) for disks with viscosity parameter
$\alpha=10^{-4}$. Our fiducial case adops a minimum pebble radius of
$a_{p,{\rm min}}=0.01\:$cm (left column panels of
Figure~\ref{fig:peb_flux1e-4}), but we also show the results for
$a_{p,{\rm min}}=0.1\:$cm (right column). Our fiducial choice of
initial and injected pebble to dust mass surface densities is
$\Sigma_p/\Sigma_d=0.01$ (middle row), but we also show the results
for ratios that are three times smaller (top row) and larger (bottom
row). Our fiducial case has a steady accretion rate of
$\dot{m}=10^{-9}\:M_\odot\:{\rm yr}^{-1}$ (black/grey lines), but we
explore results of varying this from $10^{-10}\:M_\odot\:{\rm
  yr}^{-1}$ (red lines) to $10^{-8}\:M_\odot\:{\rm yr}^{-1}$ (blue
lines).  We show the results of the case of pebble evolution where
only sweep-up growth of dust is considered (solid lines), and then
also for the case where pebble-pebble coagulation is included (dashed
lines). Finally, our fiducial model involves the initial disk from 0.1
to 30~AU being populated with pebbles with the given ratio of
$\Sigma_p/\Sigma_d$ (dark lines), however we also consider the case of
an initially ``empty'' disk, i.e., no pebbles, only dust, with pebbles
only appearing via injection at 30~AU (lighter shaded lines).  The
steady-state pebble supply rates assuming all solids are in pebbles
are shown by the horizontal dotted lines in the panels, i.e., three
lines for the three accretion rates of $10^{-10}, 10^{-9},
10^{-8}\:M_\odot\:{\rm yr}^{-1}$.

Considering the fiducial case in the left-middle panel, we see there
is an initial period of pebble supply rates that are greater than the
steady state rate due to delivery of solid material (mostly dust) that
was initialized to be present in the disk, i.e., from 0.1 to
30~AU. Pebble-pebble coagulation leads to more efficient sweep-up of
this material so that the period of elevated supply rate is shorter,
but more intense. The late-time behavior of the models is asymptotic
approach to the steady state supply rate (from above).

This trend is basically repeated in the higher $\dot{m}$
case. However, in the lower $\dot{m}$ case we see that the pebble
supply rate drops below the expected steady state rate, and then
slowly approaches it from below. Figure~\ref{fig:peb_flux1e-4}b shows the
same models, but now for $a_{p,{\rm min}}=0.1\:$cm, which has the
effect of initializing with and continuing to inject with a smaller
number of larger pebbles. Similar behavior is seen, with an initial
``spike'' phase of elevated pebble supple rates, which are shorter,
but more intense, if pebble-pebble coagulation is modeled. Now,
however, the $\dot{m}=10^{-9}\:M_\odot\:{\rm yr}^{-1}$ case also
exhibits a pebble supply rate that falls below the steady state supply
rate immediately after the spike phase.

The above behavior can be better understood, by comparing to the same
results in lighter colored lines that show the models in which there
are no initial pebbles in the disk: they only appear via injection at
the 30~AU boundary. These models show initial elevated pebble supply
rates due to delivery of the initial dust from the disk, but do not
show the decremented phases where pebble accretion rate is below the
steady state rate.

We conclude that the initial spike phase, due to delivery of initial
disk dust, is a somewhat artificial feature of our model, dependent on
the initial conditions. A subsequent decremented supply rate phase can
occur in some cases if there has been very efficient sweep-up of the
initial dust by an initial population of pebbles and there is a
relatively long period of time needed for the injected pebbles to
drift in to the inner disk. The limiting case of no initial pebbles in
the disk is instructive in that it tells us the time needed for the
disk to reach a quasi equilibrium state where pebble supply rate at
the inner boundary equals that injected at 30~AU. For only sweep-up
growth, this timescale can be $\gtrsim1\:$Myr for the high accretion
rate case, and somewhat shorter for lower $\dot{m}$ cases. However,
including pebble-pebble coagulation reduces these timescales. Since
early high accretion rate phases are expected to last
$\lesssim1\:$Myr, i.e., the evolution from the protostellar disk
phase, this tells us that a globally self-consistent model may need to
take into account the evolving disk structure due to a declining
accretion rate and that a steady state may not be reached in certain
circumstances.

\begin{figure*}[!h]
\centering
\includegraphics[width=1.0\textwidth]{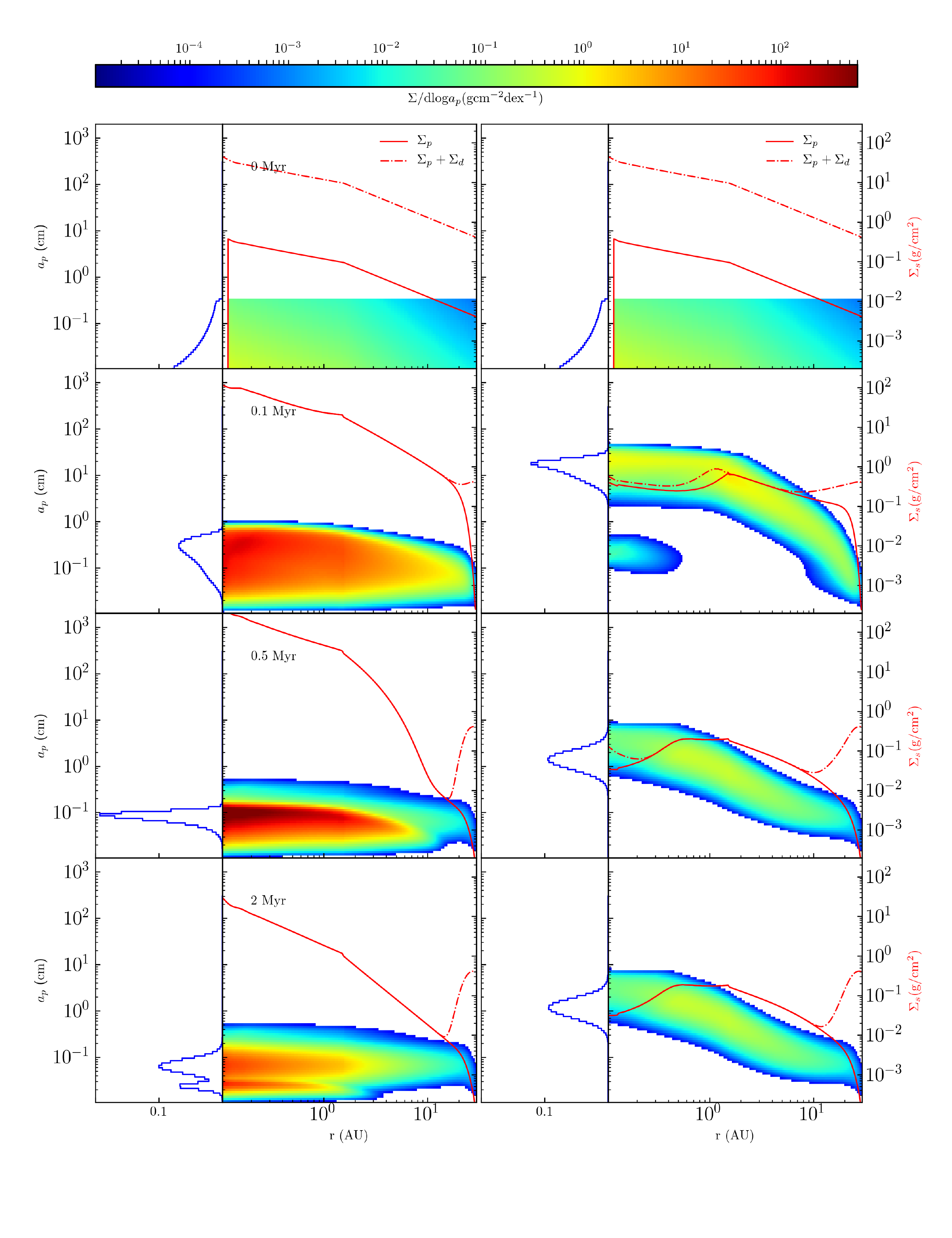}
\vspace{-0.7in}
\caption{
Evolution of pebble size versus orbital radius distributions from the
initial condition (top row), to 0.1~Myr (2nd row), to 0.5~Myr (3rd
row) to 2~Myr (bottom row). The left column shows the sweep-up only
growth model. The right column shows the sweep-up plus pebble-pebble
coagulation growth model. In each panel the color scale shows the
distribution of the pebbles, while on the left there is a projected
histogram of the sizes of those pebble inside 1~AU radii. The red
solid line shows the mass surface density of pebbles. The red
dot-dashed line shows the total mass surface density of solids (dust
plus pebbles). When these overlap, most of the solid mass is in
pebbles, which is achieved by late times in both models. 
}
\label{fig:dens2d}
\centering
\end{figure*}

Figure~\ref{fig:dens2d} shows the radial profiles of the dust size
distributions for our fiducial model with and without pebble-pebble
coagulation. The size distribution of pebbles delivered to the inner
disk is significantly larger in the latter case. The figure also shows
the radial profiles of the mass surface densities of pebbles and total
solids, which allows assessment of the mass fraction of solids that is
in pebbles compared to dust (recall that our modeling is of the
midplane region).
 
In models with pebble coagulation, the sweeping up of dust is not as
efficient in the inner disk as in the sweep-up only model. As seen
from Figure~\ref{fig:dens2d}, the pebble mass surface density is
significantly lower than the total solids mass surface density, while
in the sweep-up only model, pebbles dominate the total solid mass
starting even at 0.1~Myr. The growth rate of pebble coagulation scales
with the square of pebble number density $N_p^2$, while sweep up
growth scales with $N_p$.  During the early stages in the inner disk,
where the pebble density is high, fast mutual coagulation produces
large, Stokes-limited pebbles without sweeping up much dust. The inner
disk can maintain a high dust fraction until the pebble flux from the
outer disk arrives.

Another feature worth noticing in the coagulation model is the
transient behavior at 0.1~Myr. The pebbles are divided into two groups
in the inner disk. There are two contributing factors: the
Stokes-limited sweep-up growth forbids the further sweep-up growth
of some initial pebbles in the inner disk; and the 15 m/s coagulation
speed limit stops them being coagulated into larger sizes, while
their mutual coagulations are very inefficient due to small cross
sections and low relative velocities (see Figure~\ref{fig:delta_v}).

\subsection{Effects of $\alpha$ Viscosity}

\begin{figure*}[t]
\centering
\includegraphics[width=1.0\textwidth]{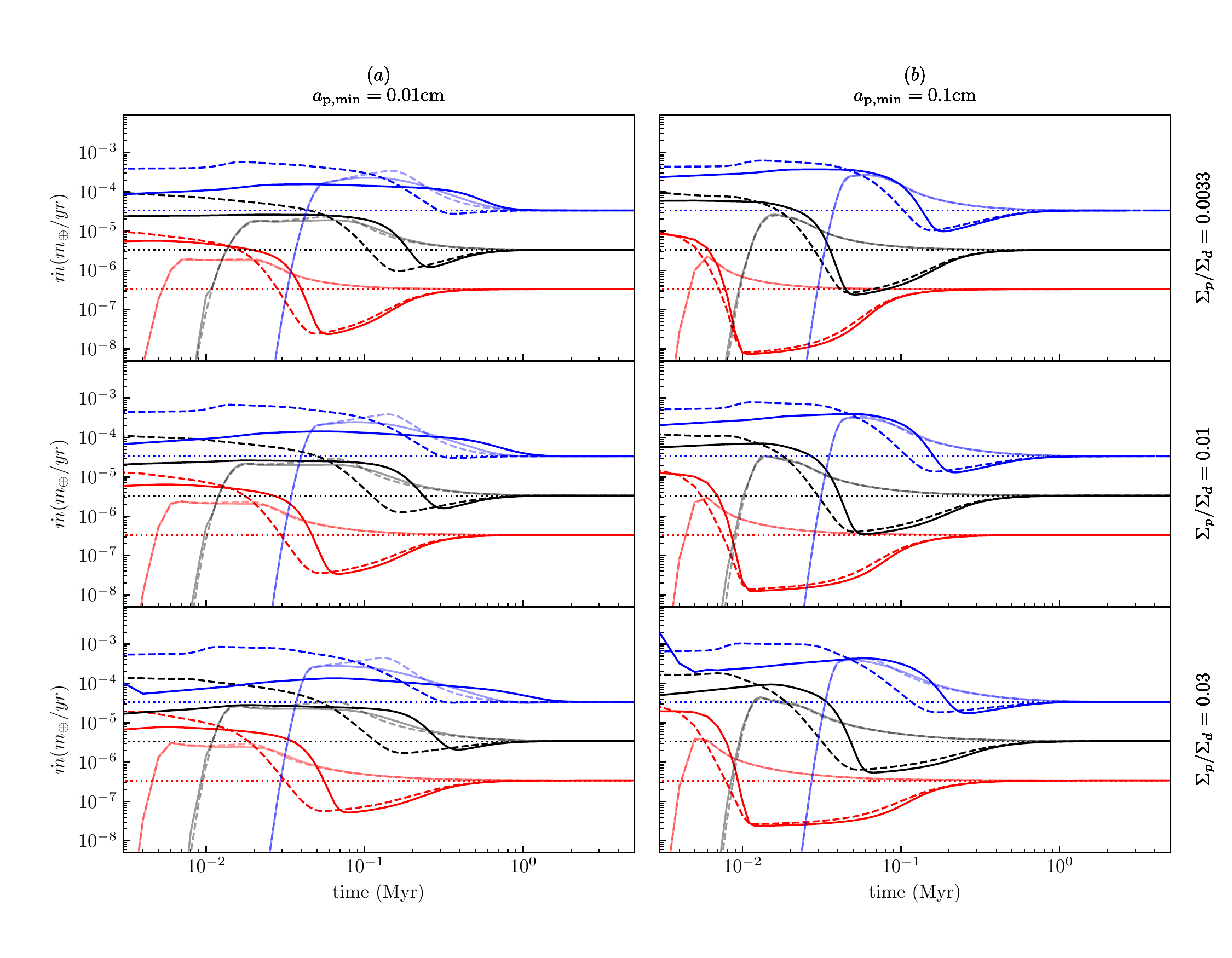}\\
\caption{
{\it (a) Left:} As Figure~\ref{fig:peb_flux1e-4}a, but now for disks with
$\alpha=10^{-3}$.%
{\it (b) Right:} As (a) but now for $a_{p,{\rm min}}=0.1\:$cm.}
\label{fig:peb_flux1e-3}
\centering
\end{figure*}

\begin{figure*}[!h]
\centering
\includegraphics[width=1.0\textwidth]{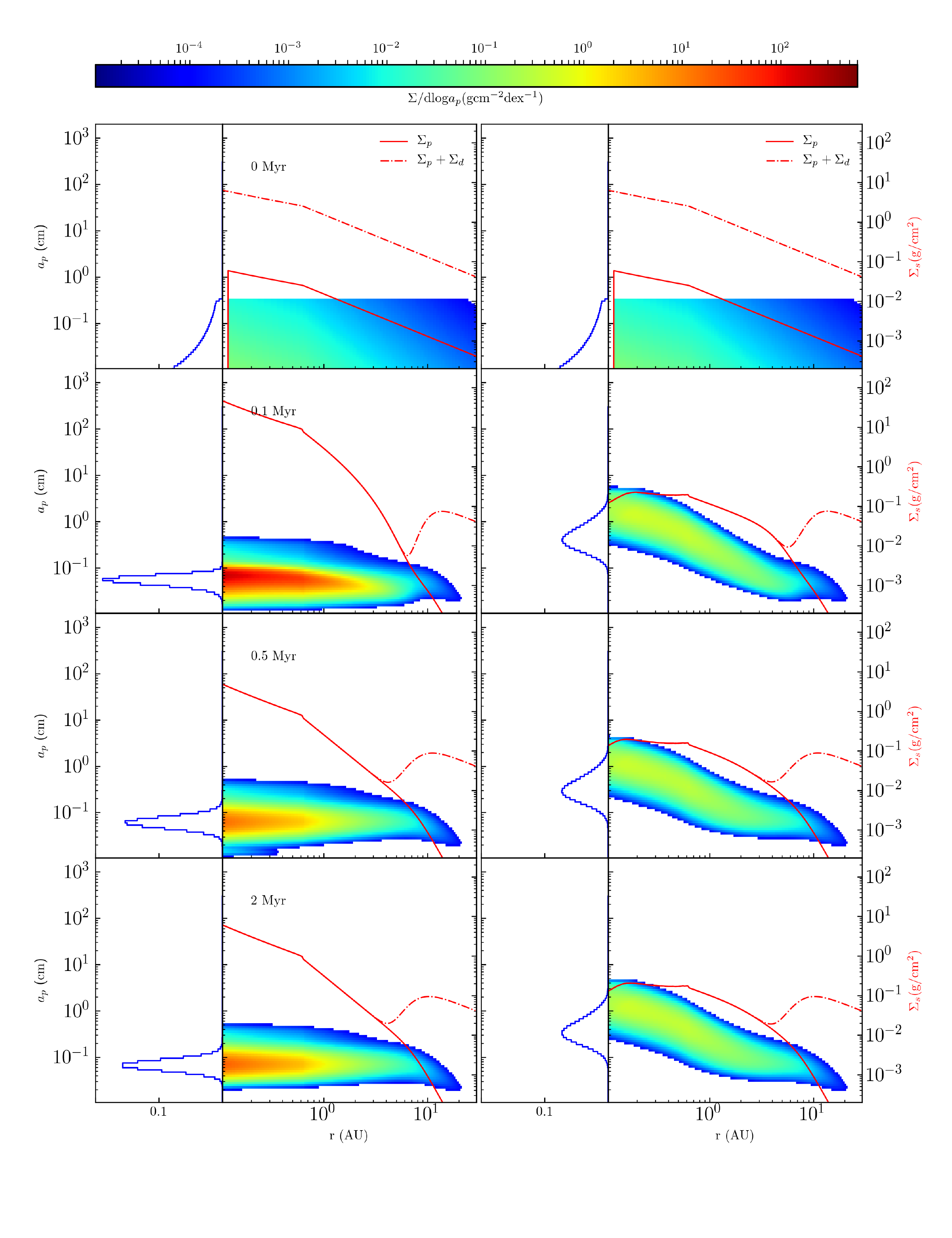}
\vspace{-0.7in}
\caption{
Same as Figure~\ref{fig:dens2d}, but showing disk with viscosity
parameter $\alpha=10^{-3}$. The major difference is in the outer disk:
significantly more solids remain in the form of dust than in the disk
with $\alpha=10^{-4}$, thus resulting in a lower pebble flux after the
dust in the inner disk has been swept-up.
}
\label{fig:dens2d-3}
\centering
\end{figure*}

The appropriate value of the effective viscosity, parameterized via
$\alpha$, is very uncertain. Here we repeat the analysis of the
previous sub-section, but now for disks with $\alpha=10^{-3}$.  These
results are shown in Figure~\ref{fig:peb_flux1e-3}.  In our modeling,
variation of $\alpha$ plays a similar role as $\dot{m}$, as it mainly
affects mass surface density, i.e., $\Sigma_g \propto \alpha^{-4/5}$
(see eq.~\ref{eq:sigma_g}). Higher $\alpha$ disks thus have a smaller
initial dust reservoir, i.e., in the scale of the 30~AU disk, and less
efficient sweep-up growth. Thus, note that the pebble flux profile in
a disk of $\dot{m}=10^{-9}\:M_\odot\:{\rm yr}^{-1}, \alpha=10^{-3}$ is
quite similar to the one with $\dot{m}=10^{-10}\:M_\odot\:{\rm
  yr}^{-1}, \alpha=10^{-4}$. Though the dust in the inner disk
transfers mostly into pebbles, as in the low $\alpha$ case, the solids
in the outer disk stay mostly in the form of dust for a much longer
time. This effect is visible in Figure~\ref{fig:dens2d-3}, which
shows the radial dependence of the pebble size distributions and the
relative levels of $\Sigma_p$ and $(\Sigma_p+\Sigma_d)$ for the higher
$\alpha$ disks.

\section{Assembling a STIP}\label{S:STIP}

\subsection{Steady Accretion Rate Disks}

\begin{figure*}[h]
\centering
\includegraphics[width=1.0\textwidth]{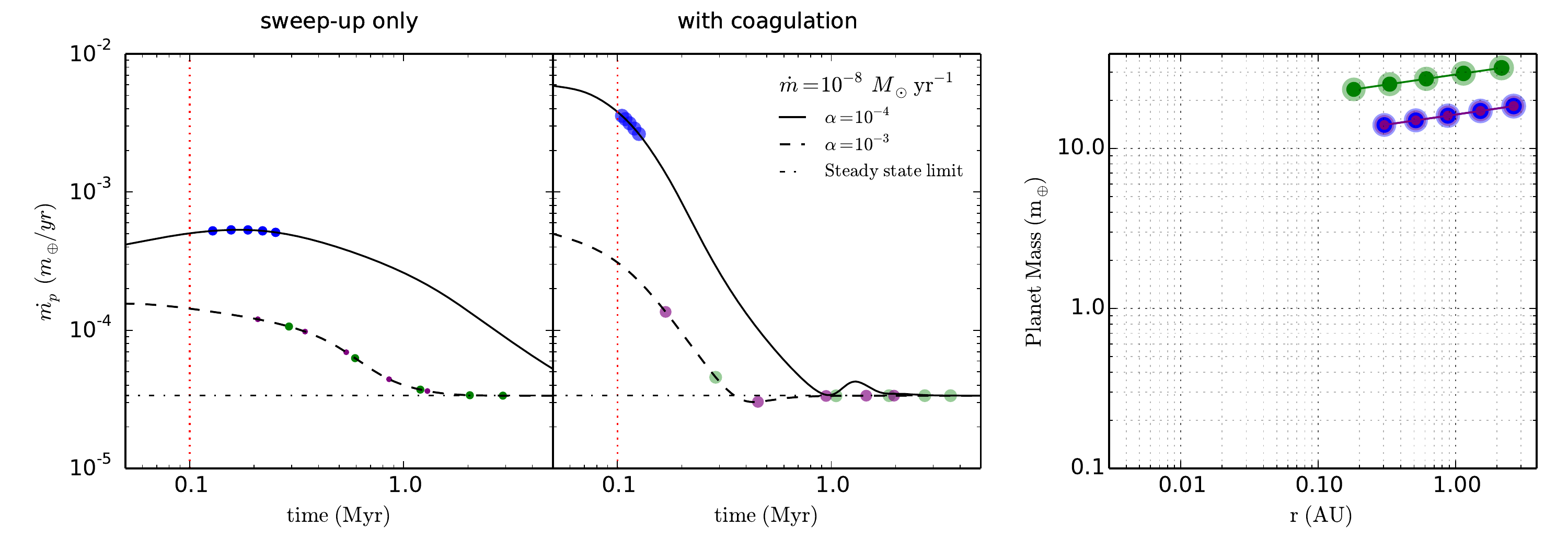}
\includegraphics[width=1.0\textwidth]{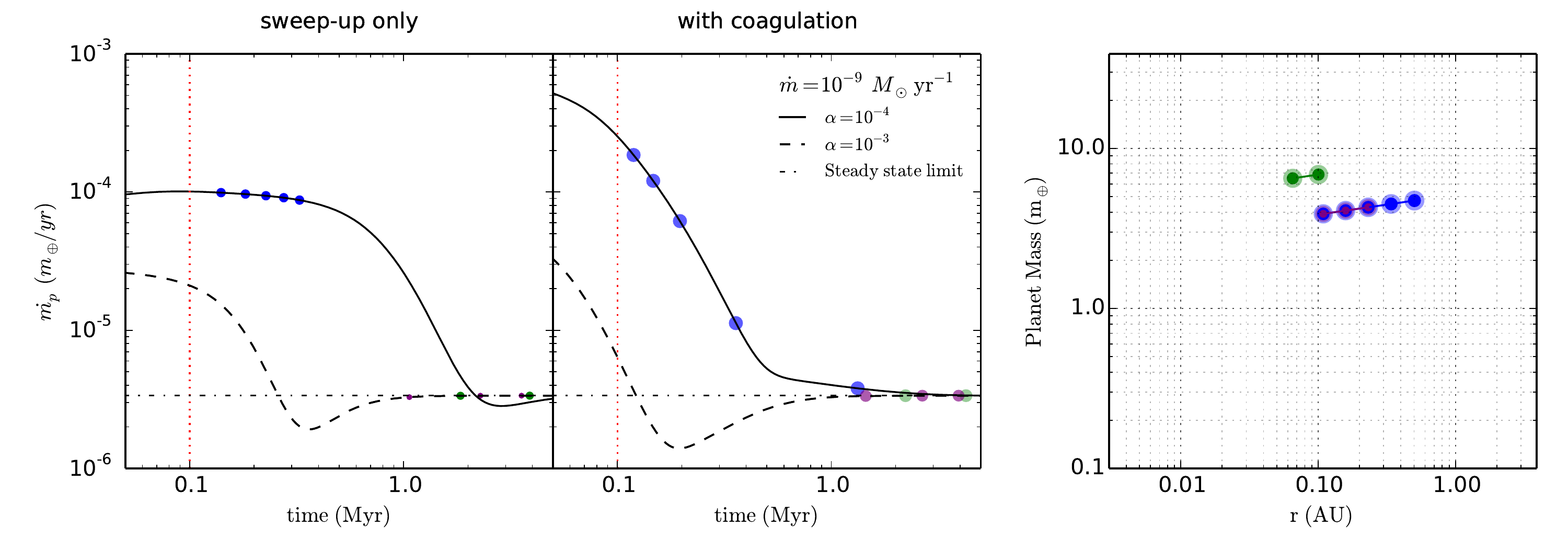}
\includegraphics[width=1.0\textwidth]{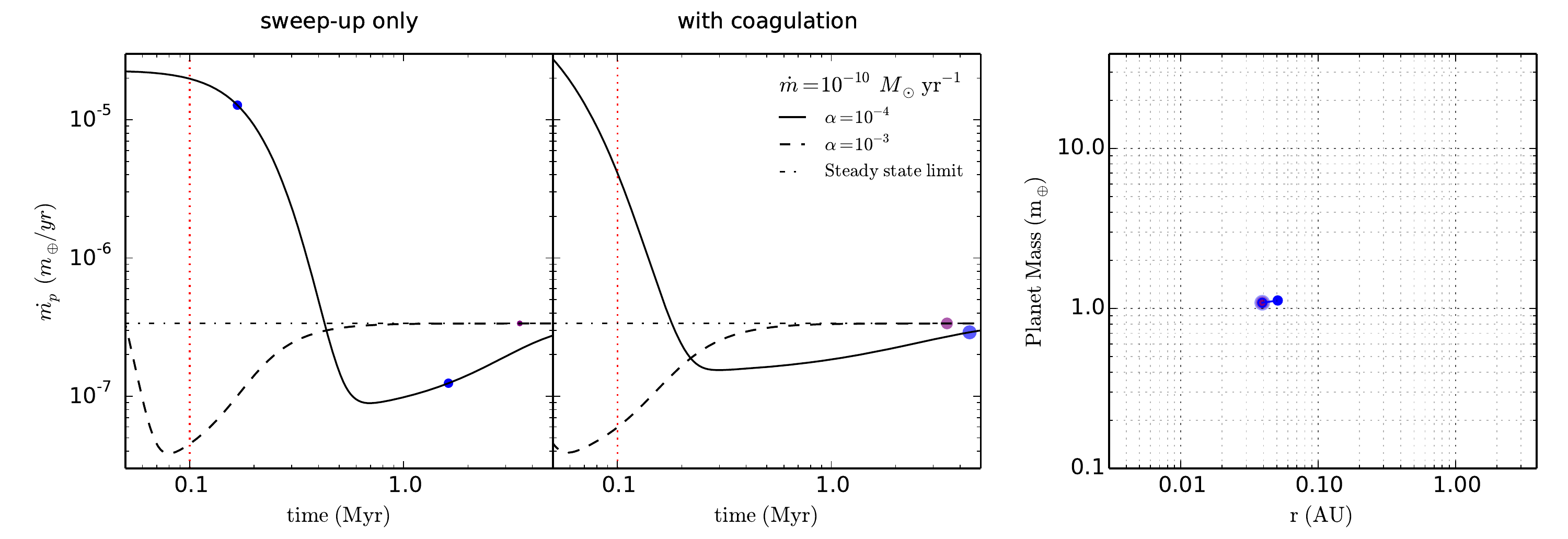}
\caption{
{\it (a) Top row:} Example of STIP formation history (left and middle
panels) and orbital architecture, i.e., planet mass versus orbital
radius (right panel), given a pebble supply rate history for a model
with $a_{p,{\rm min}}=0.01\:$cm and a steady gas accretion rate of
$\dot{m}=10^{-8}\:M_\odot\:{\rm yr}^{-1}$.  The results of models with
only sweep-up growth of pebbles are shown in the left panel. The black
solid line shows the pebble supply rate to the inner disk with
$\alpha=10^{-4}$, and black dashed line shows pebble supply rate with
$\alpha=10^{-3}$. The dot-dashed line marks the steady state pebble
flux limit. The colored dots on the pebble delivery lines indicate the
formation time of each planet, with blue dots for $\alpha=10^{-4}$,
green dots for $\alpha=10^{-3}$, and purple dots for global disk
$\alpha=10^{-3}$ but with calculation of $M_{G,D}$ done for
$\alpha=10^{-4}$, i.e., a lower value of $\alpha$ near the DZIB. Note
that in these models planet formation is assumed to start at 0.1~Myr
(marked by the vertical dashed line). The middle panel shows the
equivalent results, but now including pebble-pebble coagulation.
 {\it (b) Middle row:} As (a) but for disk accretion rates of
 $\dot{m}=10^{-9}M_\odot\:{\rm yr}^{-1}$.
{\it (c) Bottom row:} As (a) but for disk accretion rates of
$\dot{m}=10^{-10}M_\odot\:{\rm yr}^{-1}$. The disks with lower
accretion rates can fail to form all five planets within the 5 Myr
period considered. 
 }
\label{fig:mdot_planet}
\centering
\end{figure*}

We now consider the implications of these models for the timescale of
assembling an entire STIP. As a specific example, we will consider the
ability of disk models to form a five-planet system and the time it
takes them to do so. We compare the accumulated mass from the
delivered inner disk pebble flux to the required mass of the planets,
i.e., set by the gap opening condition for $M_{\rm G,D}$ from
eq.~\ref{eq:mg1}.

A noticeable feature of our modeled pebble flux is the ``spike'' phase
at the beginning of the simulations, which can transport a large
amount of solids ($\sim 50 M_\oplus$) in less than 0.5~Myr. However,
this spike feature mostly reflects the adopted initial condition of
pebbles and dust in the inner ($\lesssim10\:$AU) disk. Since these
solids in reality could have already drifted in at an earlier phase of
the disk's evolution, here we ignore the pebbles delivered in the
early spike phase up to 0.1~Myr. Still, we will see that our results
do depend on how the initial conditions of the disks are set-up.

In the context of IOPF, building a STIP starts with the innermost
``Vulcan'' planet at $r_{\rm 1200K}$. After that, here, for
  simplicity, we assume that the orbit of each subsequent planet is
20 Hill radii away from the previously formed planet, which is close
to the median value from observations (Paper I) and within
  the range of DZIB retreat distances modeled in Paper III as being
  due to increased ionization via penetration of protostellar
  X-rays.  The planet mass is calculated as $M_{\rm G,D,1.44}$ using
eq.~\ref{eq:mg1}. We simply assume all accumulated mass of pebbles
will be transferred into planets, i.e., the radially drifting pebbles
will be caught by the DZIB pressure trap and later accreted with
  100\% efficiency by the protoplanet.

The results of this modeling for our fiducial disk ($\dot{m}=10^{-9}
\:M_\odot\:{\rm yr^{-1}}, \alpha=1\times10^{-4}$) are shown with solid
lines in the left pair of panels in the middle row of
Figure~\ref{fig:mdot_planet}. Here, the first planet starts forming
0.1~Myr after the spike phase, and the fifth has completed formation
within the following 0.4~Myr in the case of sweep-up only pebble
evolution and within 1.5~Myr in the case including pebble-pebble
coagulation (blue dots mark the time of formation of each planet). The
difference is due the different pebble supply rates, with the first
model more heavily influenced by the initial spike phase from the
initial dust reservoir. The orbital distribution and masses of the
planets are shown in the rightmost panel of the middle row (again
shown as blue dots).

Equivalent models with $\alpha=1\times10^{-3}$ tend to have lower
pebble supply rates (dashed lines). If this value of $\alpha$ also
holds in the DZIB region, then the gap opening mass is increased,
implying more massive planets. In this case only two such planets
(green dots) have time to form within 5~Myr. 

The top and bottom rows of Figure~\ref{fig:mdot_planet} show
equivalent results for $\dot{m}=10^{-8}\:M_\odot\:{\rm yr^{-1}}$ and
$\dot{m}=10^{-10}\:M_\odot\:{\rm yr^{-1}}$, respectively. Higher/lower
accretion rates imply higher/lower gap-opening masses, but with a
sub-linear scaling. The total mass of pebbles delivered to the inner
region in a given time scales more steeply with accretion rate. Thus,
it becomes more difficult for lower accretion rate disk models to be
able to form the entire five planet system. Overall, when comparing
models with different $\dot{m}$ and $\alpha$, we find that to form a
realistic STIP, the disk should have $\dot{m}\sim 10^{-9}
M_\odot\:{\rm yr}^{-1}$ and viscous $\alpha$ no more than
$\sim2\times10^{-4}$.

\subsection{Evolving Disks}
\label{S:evo}

Lifetimes of protoplanetary disks are estimated to range from $\sim$1
to 10 Myrs \citep[e.g.,][]{2014prpl.conf..475A, 2014prpl.conf..497E}.
The effects of an evolving disk
have been previously considered in dust evolution models. For example,
\citet{Birnstiel2010} set up a disk model subject to viscous evolution
and gravitational instability, starting from the very early infall
phase. It gives a relatively constant accretion rate starting from
0.15~Myr to 1~Myr. Other methods involve adopting simple functional
forms for the evolution of the accretion rate, such a power law or
exponential decay \citep[e.g.,][]{HG2005,2015A&A...582A.112B}. The
disk modeled by \citet{2015A&A...582A.112B} has an accretion rate that
evolves as:
\begin{eqnarray}
\dot{m}= \dot{m}_0 e^{-t/t_0},
\end{eqnarray}
with $\dot{m}_0=7.24\times10^{-9}\:M_\odot\:{\rm yr^{-1}}$ and
$t_0=3.1\times10^5\:$yr. We adopt the same functional form, but with
$\dot{m}_0=1.0\times10^{-9}\:M_\odot\:{\rm yr^{-1}}$ and
$t_0=4.34\times10^5\:$yr, so that the accretion rate declines by a
factor of 10 in 1~Myr.

In our numerical modeling, we update the disk properties, i.e., the
accretion rate, every 1000 years.  Note, in such models of
exponentially declining accretion rate, the disk evolution is not
fully self consistent for its accretion history, i.e., the declining
$\Sigma_g$ does not match the viscous radial drift of gas. Thus these
models should be viewed as being simple ``toy'' models of disk
evolution. Note also, when we force the gas disk properties to evolve
with an exponentially declining rate, we have a choice about how we
treat the associated dust that may be mixed together with the gas.  We
consider two extreme cases: {\it Case A:} the dust content of the
disk, i.e., $\Sigma_d$, does not evolve with the declining gas, i.e.,
it would remain at a steady mass surface density, except for the
effects of radial drift (due to the assumed $\alpha$ viscosity) and
sweep-up by pebbles; {\it Case B}, the dust content of the disk, i.e.,
$\Sigma_d$, is decreased with the same exponential decay rate of the
gas, with this decay factor applied multiplicatively on top of any
other changes due to radial drift and sweep-up by pebbles.

In principle, there are now several parameters that may be varied in
these models: $\dot{m}_0$, $t_0$, $t_{\rm IOPF}$ (i.e., the time when
the first planet starts to form), in addition to other model
parameters already introduced, such as $\alpha$ and $a_{p,{\rm
    min}}$. Furthermore, there are the choices of: Case A or B,
introduced above; whether or not pebble-pebble coagulation is included
in the modeling; and whether or not the disk model is initially
``empty'' of pebbles. It is not our intention here to carry out a full
exploration of the parameter space of all these models. Rather, here
we present several example models of representative cases, in
particularly focussing on the examples of parameter combinations that
produce realistic STIPs within reasonable timescales.

\begin{figure*}[h]
\centering
\includegraphics[width=1.0\textwidth]{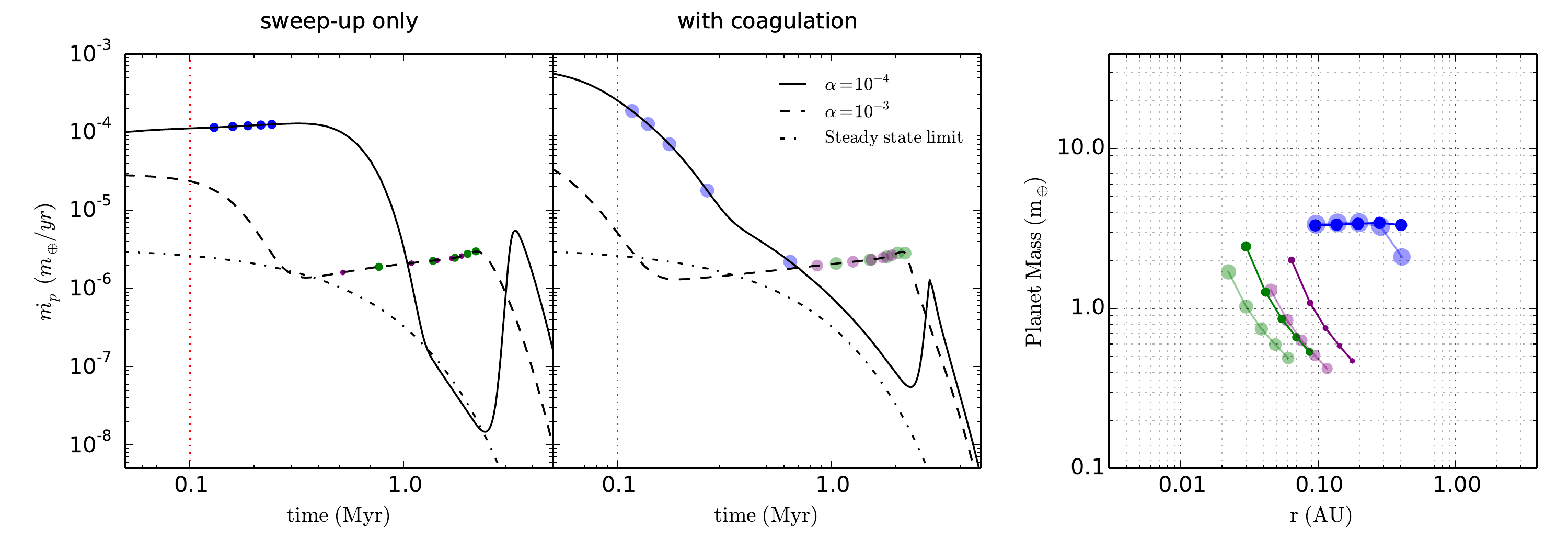}
\includegraphics[width=1.0\textwidth]{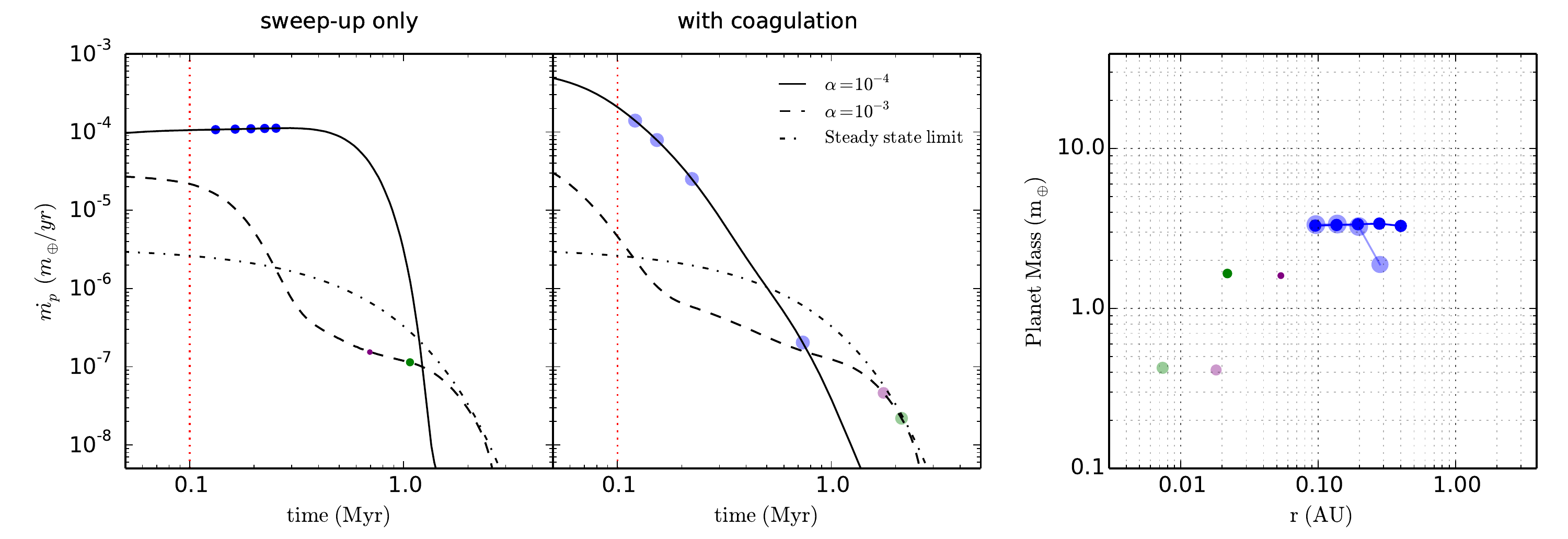}
\caption{
{\it (a) Top row:} Example of STIP formation history (left and middle
panel) and orbital architecture (right panel) in an evolving disk
(treated as {\it Case A}, see text), using same color coding and
styles as Figure~\ref{fig:mdot_planet}. Here, planet formation also
starts at $t_{\rm IOPF}=0.1\:$Myr. The only addition is the lighter
shaded points in the right panel represent planets formed in the
``with coagulation'' model.
{\it (b) Bottom row:} As (a), but now for {\it Case B}, i.e., dust
content of disk is also scaled with the exponential decay function.
}
\label{fig:evo_planet}
\centering
\end{figure*}
\begin{figure*}[h]
\centering
\includegraphics[width=1.0\textwidth]{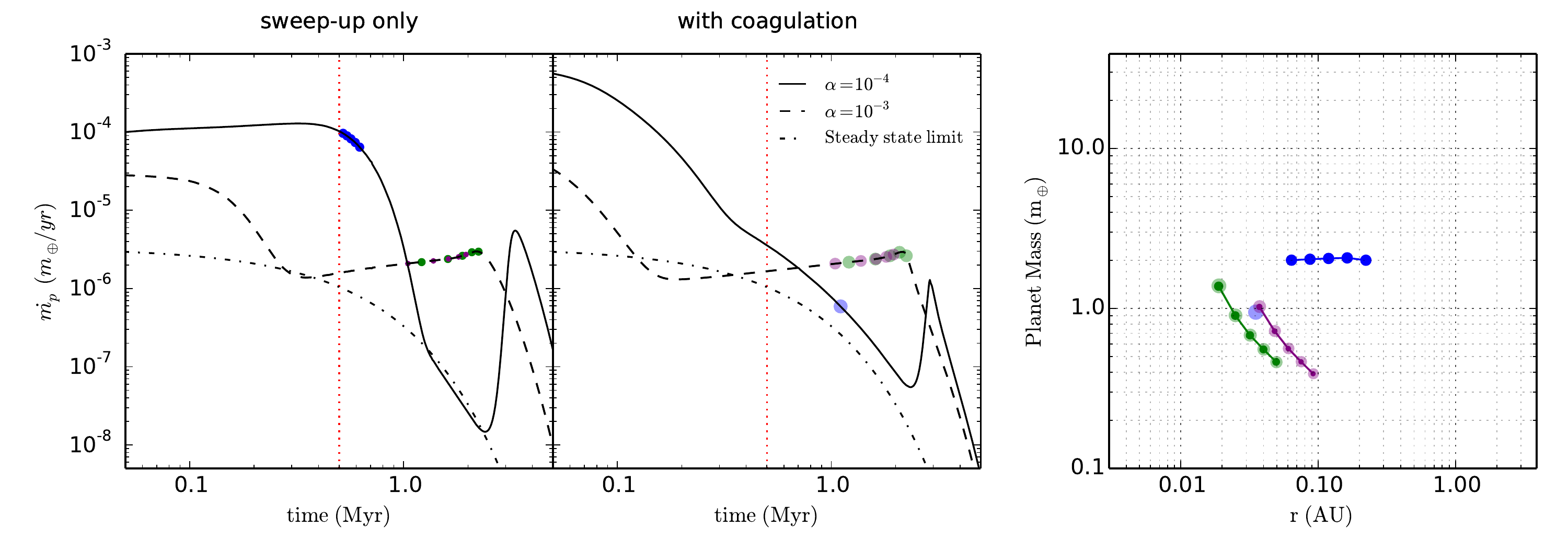}
\includegraphics[width=1.0\textwidth]{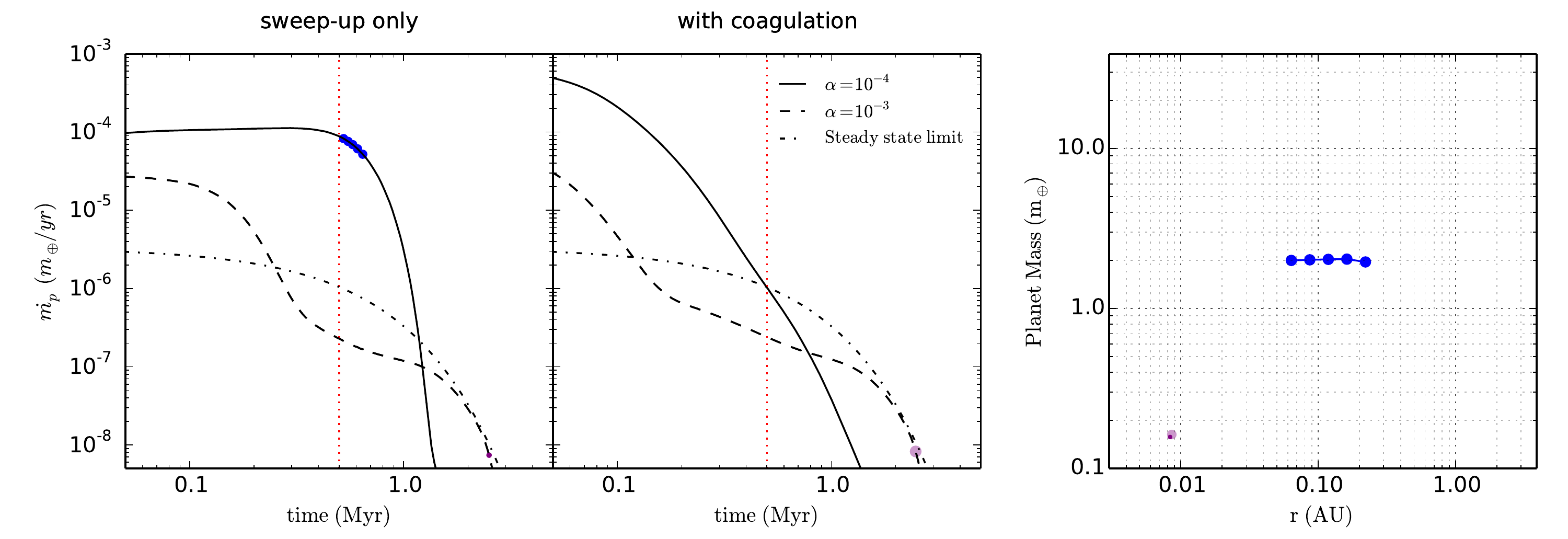}
\caption{
{\it (a) Top row:} Same as Figure~\ref{fig:evo_planet}a, i.e., {\it
  Case A} models, but now with planet formation starting at $t_{\rm
  IOPF}=0.5\:$Myr.
{\it (b) Bottom row:} As (a), but now for {\it Case B}, i.e., dust
content of disk is also scaled with the exponential decay function.
}
\label{fig:evo_planet2}
\centering
\end{figure*}

Following the format of Figure~\ref{fig:mdot_planet},
Figure~\ref{fig:evo_planet} now shows example models with
$\dot{m}_0=10^{-9}\:M_\odot\:{\rm yr}^{-1}$, $t_0=4.34\times10^5\:$yr,
$t_{\rm IOPF}=0.1\:$Myr and $a_{p, {\rm min}}=0.01\:$cm. Case A is
shown in the top row; Case B in the bottom row. In each row, models
with low $\alpha$ ($10^{-4}$), high $\alpha$ ($10^{-3}$) and ``mixed''
$\alpha$ (i.e., $10^{-4}$ in the DZIB region; $10^{-3}$ in the
  main disk region beyond the DZIB region that is relevant for pebble
  growth and supply) for the cases without and with pebble-pebble
coagulation. In these models, the value of $\alpha$ in the DZIB
  region has a direct effect on the mass of the planets needed to open
  up a gap, i.e., lower mass planets in lower viscosity disks. The
  value of $\alpha$ in the main disk region beyond the DZIB affects
  global disk structure in the zone responsible for pebble growth and
  supply: in particular, lower viscosity implies a higher disk mass,
  for a given accretion rate, and so a more massive reservoir of dust
  for planet formation. The ``mixed'' $\alpha$ model that is presented
  is a simple means of allowing for there being different values of
  viscosity in DZIB that affect planet masses and in the main disk
  region that affect pebble growth and supply.

Compared to the steady disk models with
$\dot{m}=10^{-9}\:M_\odot\:{\rm yr}^{-1}$, these evolving disk models
have lower late-time pebble delivery rates. Planets forming at these
later times have lower masses, since the gap opening mass is lower for
lower accretion rates. We see that certain combinations of models lead
to planet formation being spread out in time over a few Myr, which
leads to a significant decline in planet mass with orbital
radius. This is not a particular feature of observed STIPs, modulo
uncertainties in inferring their masses from observed sizes (see,
e.g., Paper I, II). To obtain relatively flat distributions of planet
masses versus orbital radii requires IOPF to be mostly complete within
$\sim t_0$ (and thus for $t_{\rm IOPF}\ll t_0$). Only the low $\alpha$
models achieve these conditions. 

The influence of the spike phase of enhanced pebble supply rate at
early times can extend the time period when reasonable STIPs can be
formed by IOPF, especially in the sweep-up only pebble evolution
models. This is also illustrated in Figure~\ref{fig:evo_planet2},
which shows the same results as in Figure~\ref{fig:evo_planet}, but
now for $t_{\rm IOPF}=0.5\:$Myr. Only in the low $\alpha$, sweep-up
only models does a STIP with flat mass versus orbital radius
distribution form. 

We note that the flux increase after about 2.5 Myr in {\it Case A} is
a result of both leftover dust from early phase and the relatively
large Stokes numbers of the injected pebbles at late stages. The
injected pebbles, i.e., with radii of 0.01 cm to 0.3 cm, have
relatively small Stokes numbers in the early phases, when gas
densities are higher. However, the Stokes number increases
significantly as $\dot{m}$ drops. Still, in the models considered
here, this late-time pulse of delivered pebbles is not significant
compared to the early pebble delivery that formed the STIP.

Note these are systems with modest initial accretion rates: $\dot{m}=
10^{-9} M_\odot\:{\rm yr^{-1}}$. If IOPF starts at even higher
$\dot{m}$, then the planets formed would be significantly more massive
than typical observed STIPs (unless the DZIB $\alpha$ is much smaller
than values we have considered).

\section{Discussion and Conclusions}\label{S:conclusions}

In this paper we have presented improved models of Inside-Out Planet
Formation (IOPF), with a particular focus on understanding if most
solids will be delivered to the inner disk in the form of pebbles
(i.e., the raw materials for planet formation) and then the implied
formation times and constraints on disk properties during the time of
IOPF. The models presented here are an attempt at calculating ``IOPF
population synthesis'', and can be compared to other such models based
on other theoretical frameworks
\citep[e.g.,][]{2004ApJ...604..388I,2015A&A...582A.112B}.

We first constructed improved disk models, extending to 30~AU,
especially using realistic opacities and including the effect of the
active disk to passive disk transition. Next, we expanded on the work
of Paper III to examine the gap opening condition (i.e., pressure
maximum displacement from the planet) over a wider range of
viscosities (especially extending to the low $\alpha$ regime that
seems to be needed for IOPF) and accretion rates. We find that the
\citet{Duffell2015} analytic description of the gap opening criterion
is a better description of the results than the simple viscous
criterion we used previously in Papers I to III. This implies that
planets are more massive in the low $\alpha$ regime, i.e., still $\sim
4\:M_\oplus$ for $\alpha=10^{-4}$ for $\dot{m}=10^{-9}\:M_\odot\:{\rm
  yr}^{-1}$.

Next we examined pebble evolution in these disks, presenting models
for how a single pebble is expected to evolve via Stokes-limited
sweep-up of dust and pebble-pebble coagulation. This formed the basis
of a global pebble evolution model, i.e., of a population of pebbles,
for given choices of initial conditions and boundary conditions (i.e.,
the injected pebbles at 30~AU).

Simple constraints on timescales of first, ``Vulcan'' planet formation
were derived with the single pebble evolution model, under the
assumption of a given efficiency of solids being incorporated into
pebbles (Table~\ref{tab:zone}). Fiducial estimates based on drift
times from the reservoir radius are $\sim 10^5\:$yr to form the first
planet. The global pebble evolution models showed that large, near
unity fractions of solids are expected to be incorporated into pebbles
(with the caveat that our disk models apply to the midplane layer
conditions; but this is where most solids are expected to have
settled). By the time they reach the inner AU, pebble sizes are
typically $\sim0.1\:$cm in the sweep-up only models and about a few cm
in the models including pebble-pebble coagulation. Such predictions
are potentially testable by observations of protoplanetary disks
\citep[e.g.,][]{2012ApJ...760L..17P,2013A&A...558A..64T,
  2014prpl.conf..339T,2016A&A...588A..53T}.

We utilized the global pebble evolution models in disks with steady
accretion rates to predict pebble delivery rates to the inner disk. We
showed the effects of model assumptions, including whether the disk is
initially populated with pebbles. The models show an initial ``spike''
phase of an elevated pebble flux compared to the steady state rate due
to sweeping up of the initial dust reservoir, which decays away in
about a Myr.

Finally we used these models to predict the formation history of STIPs
forming via IOPF. Formation locations of the Vulcan planets were set
self-consistently at $r_{\rm 1200K}$ and with masses that depend on
accretion rate, location, and disk viscosity. The timescale to form a
five-planet system was assessed, with an overall cap of 5~Myr
imposed. In steady accretion rate disks, these models showed a
preference for $\dot{m}\sim10^{-9}\:M_\odot\:{\rm yr}^{-1}$. Higher
accretion rate models form planets that are too massive and too far
away from the star. Lower accretion rate models do not have high
enough pebble delivery rates to form several planets. Extending to
evolving disk models introduces additional parameters and model
choices. For simple evolving disk models treated with an exponentially
declining accretion rate, we require IOPF to be completed within this
decay time $\sim 1\:$Myr in order to preserve the observed relatively
flat scaling of planet mass with orbital radius.

We note that IOPF's apparent preference for accretion rates of
  $\dot{m}\sim10^{-9}\:M_\odot\:{\rm yr}^{-1}$ is quite consistent
  with observed accretion rates of young stellar objects, including
  transition disk systems \citep[e.g.,][]{2011ARA&A..49...67W, manara,
    manara16}. At even lower accretion rates, the rate of
  photoevaporation of the disk may become comparable, potentially
  terminating or greatly reducing the supply of gas to the inner disk
  \citep[e.g.,][]{ercolano2009, owen12, ercolano2014}. However, if the
  solids are mostly settled to the disk midplane in the form of
  pebbles, as indicated by the results of our global pebble evolution
  models, then their supply is less likely to be affected by
  photoevaporation. Nevertheless a more complete global evolution
  model of IOPF, planned for a future study, should also account for
  additional processes such as photoevaporation to examine its effects
  on gas supply to the inner disk region. Self-consistent constraints
  of disk lifetimes that are predicted by either residual core infall
  and/or photoevaporation will also need to be considered in this
  modeling.

\acknowledgements We thank an anonymous referee for helpful comments
that improved the clarity of the paper. JCT acknowledges NASA ATP
grant NNX15AK20G and NSF AAG grant 1616300. ZZ acknowledges NASA ATP
grant NNX17AK40G and a Sloan Research Fellowship. TB acknowledges
funding from the European Research Council (ERC) under the European
Union’s Horizon 2020 research and innovation programme under grant
agreement No 714769. ANY acknowledges NSF AAG grant 1616929. SM and
JCT acknowledge a Royal Society International Exchange grant.

\end{CJK*}
\end{document}